\documentclass{aa}
\usepackage{times,float}
\usepackage{epsfig}
\usepackage{graphics}
\usepackage{amssymb}

\begin{document}

\title{Strategies for prompt searches for GRB afterglows: the discovery  
of the  \object{GRB~001011} optical/near-infrared counterpart
using colour-colour selection\thanks{Based on observations
collected   at  the  European Southern   Observatory,   La Silla and
Paranal, Chile  (ESO Programmes  165.H--0464(A), 165.H--0464(E)  and
165.H--0464(G)).}}

\titlerunning{The Afterglow of GRB~001011}
\author{
   J. Gorosabel \inst{1,2}
    \and J.U. Fynbo \inst{2}
    \and J. Hjorth \inst{3,4}
    \and C. Wolf   \inst{5}
    \and M.I. Andersen \inst{6} 
    \and H. Pedersen \inst{3}
    \and L. Christensen \inst{3}
    \and B.L. Jensen \inst{3}
   \and P. M$\o$ller \inst{2}
   \and J. Afonso \inst{7}
   \and M. A. Treyer\inst{8}
   \and G. Mall\'en-Ornelas \inst{9}
   \and A.J. Castro-Tirado \inst{10,11}
   \and A. Fruchter \inst{12}
   \and J. Greiner \inst{13}
   \and E. Pian \inst{14}
   \and P. M. Vreeswijk \inst{15}
   \and F. Frontera \inst{16}
   \and L. Kaper \inst{15}
   \and S. Klose   \inst{17}
   \and C. Kouveliotou \inst{18}
   \and N. Masetti \inst{16}
   \and E. Palazzi \inst{16}
   \and E. Rol \inst{15}
   \and I. Salamanca \inst{15}
   \and N. Tanvir \inst{19}
   \and R.A.M.J. Wijers \inst{20}
   \and E. van den Heuvel \inst{15}
}

\offprints{J. Gorosabel}

\institute{ 
           Danish Space Research Institute,
           Juliane Maries Vej 30, DK--2100 Copenhagen \O, Denmark.
           e-mail: {\tt jgu@dsri.dk}
           \and
           European Southern Observatory,
           Karl--Schwarzschild--Stra\ss e 2,
           D--85748 Garching, Germany.
           e-mail: {\tt jfynbo@eso.org, pmoller@eso.org}
            \and
            Astronomical Observatory,
            University of Copenhagen,
            Juliane Maries Vej 30, DK--2100 Copenhagen \O, Denmark.\\
            e-mail: {\tt jens@astro.ku.dk, holger@astro.ku.dk, lise@astro.ku.dk, brian\_j@astro.ku.dk}
            \and
            Observatoire Midi-Pyr\'en\'ees (LAS), 14 avenue E. Belin,
            F--31400 Toulouse, France
            \and
            Max-Planck-Institut f{\"u}r Astronomie,
            K{\"o}nigstuhl 17,  D-69117 Heidelberg, Germany.
            e-mail: {\tt cwolf@mpia-hd.mpg.de}
            \and
            Division of Astronomy, P.O. Box 3000,
            FIN-90014 University of Oulu, Finland.
            e-mail: {\tt michael.andersen@oulu.fi}
            \and
            Blackett Laboratory, Imperial College, Prince Consort Road, London SW7 2BW, UK.\\
             e-mail: {\tt j.afonso@ic.ac.uk}
            \and
            Laboratoire d'Astronomie Spatiale, Traverse du Siphon, BP8,
            13376  Marseille, France.\\
            e-mail: {\tt Marie.Treyer@astrsp-mrs.fr}
            \and
            Department of Astronomy, University of Toronto, 60 St. George
            Street, Toronto, ON, M5S 3H8, Canada.\\ e-mail: {\tt mallen@astro.utoronto.ca}
            \and
            Laboratorio de Astrof\'{\i}sica Espacial y F\'{\i}sica
            Fundamental (LAEFF-INTA), P.O. Box 50727, E-28080, Madrid, Spain.\\
            e-mail: {\tt ajct@laeff.esa.es}
            \and
            Instituto de Astrof\'{\i}sica de Andaluc\'{\i}a (IAA-CSIC),
            P.O. Box 03004, E-18080 Granada, Spain.\\ e-mail: {\tt ajct@iaa.es}
            \and
            Space Telescope Science Institute, 3700 San Martin Drive, Baltimore, MD 21218, USA.
            e-mail: {\tt fruchter@stsci.edu}
            \and
            Astrophysikalisches Institut, Potsdam, Germany.
            e-mail: {\tt jgreiner@aip.de}
            \and
            Osservatorio Astronomico di Trieste, Via Tiepolo 11,
            I-34131 Trieste, Italy. e-mail: {\tt  pian@tesre.bo.cnr.it}
            \and
            University of Amsterdam,
            Kruislaan 403, 1098 SJ Amsterdam, The Netherlands. e-mail:
            {\tt  pmv@astro.uva.nl, lexk@astro.uva.nl, evert@astro.uva.nl,isabel@astro.uva.nl,edvdh@astro.uva.nl}
            \and
            Istituto Tecnologie e Studio Radiazioni Extraterrestri,
            CNR, Via Gobetti 101, 40129 Bologna, Italy. \\ e-mail: {\tt  filippo@tesre.bo.cnr.it, masetti@tesre.bo.cnr.it, eliana@tesre.bo.cnr.it}
            \and
            Th\"uringer Landessternwarte Tautenburg, D-07778 Tautenburg, Germany.
            e-mail: {\tt klose@tls-tautenburg.de}
            \and
            NASA MSFC, SD-50, Huntsville, AL 35812, USA.  e-mail: {\tt kouveliotou@eagles.msfc.nasa.gov}
            \and
            Department of Physical Sciences, University of Hertfordshire, College Lane, Hatfield,
                Herts AL10 9AB, UK. \\ e-mail: {\tt nrt@star.herts.ac.uk}
            \and
             Department of Physics and Astronomy, State University of New York, Stony Brook, NY 11794-3800, USA. \\ e-mail: {\tt rwijers@astro.sunysb.edu}
           }
%Ask johan on how to put the e-mail
%\offprints{J. Gorosabel, e-mail:{\tt  jgu@dsri.dk}}

\mail{\tt jgu@dsri.dk}

\date{Received  / Accepted }

%%%%%%%%%%%%%%%%%%%%%%%%%%%%%%%%%%%%%%%%%%%%%%%%%%%%%%%%%%%%%%%%%%%%%%%%%%%%%%%%

\abstract{  We report the discovery   of the optical and near-infrared
    counterparts to  GRB~001011.  The GRB~001011 error  box determined
    by Beppo-SAX was simultaneously imaged in the near-infrared by the
    3.58-m New  Technology Telescope and in  the optical by the 1.54-m
    Danish Telescope  $\sim8$ hr  after the gamma-ray event.  Here we
    implement  the colour-colour discrimination  technique proposed by
    Rhoads (\cite{Rhoa00a}) and extend  it using near-IR data as well.
    We present the   results provided  by an automatic   colour-colour
    discrimination   pipe-line  developed  to  discern   the different
    populations of objects present  in the GRB  001011 error box.  Our
    software  revealed three candidates  based on single-epoch images.
    Second-epoch observations   carried out $\sim$3.2 days   after the
    burst  revealed that the most   likely  candidate had faded,  thus
    identifying it with the  counterpart to the  GRB.  In deep  R-band
    images   obtained  7    months     after   the  burst   a    faint
    (R=25.38$\pm$0.25) elongated object, presumably the host galaxy of
    GRB~001011, was  detected at the position of  the  afterglow.  The
    GRB~001011 afterglow  is the first  discovered with the assistance
    of colour-colour diagram techniques.  We discuss the advantages of
    using this method and its application to error boxes determined by
    future  missions.   }

\maketitle

\section{Introduction}

Gamma-ray Bursts (GRBs)  are intense  flashes  of high energy  photons
that occur uniformly distributed on  the sky.  They were discovered in
1967 (see Bonnell \& Klebesadel  \cite{Bonn96} for a discussion of the
first  GRB  detections),  but due  to  the  lack of  rapid and precise
localisations,  their emission at other  wavelengths  was not detected
until  1997 (van    Paradijs et  al.   \cite{vanP97};  Frail   et  al.
\cite{Frai97a};  Costa et  al.   \cite{Cost97a}).  Thus,  for 30 years
they were not localised at longer wavelengths (X-ray, UV, optical, IR,
radio) and  their distances could  not be measured.   The breakthrough
that occurred in 1997 can be attributed to the advent of the Beppo-SAX
X-ray satellite (Boella et al.    \cite{Boel97a}), thanks to its  fast
(within a   few hours)  and    precise   (few arcmin  error    radius)
localisations.  Since the determination  of the redshift of GRB~970508
(Metzger  et   al.  \cite{Metz97a})  another  17 secure  spectroscopic
redshifts have been determined to  date (not considering the supernova
SN1998bw;   Galama  et al.   \cite{Gala98}),    ranging from  $z=0.43$
(Vreeswijk  et al.  \cite{Vree01};   Hjorth  et al.   (\cite{Hjor00a},
\cite{Hjor00b}) to $z=4.50$ (Andersen et  al.  \cite{Ande00}).  For an
additional 8  GRBs, optical counterparts  have been found, but with no
conclusive published redshift determinations.

It is  now   widely accepted that  at   least  the  long-duration GRBs
originate at cosmological distances (with the exception of GRB~980425)
with isotropic  equivalent energy releases  ranging  from 10$^{51}$ to
10$^{54}$ erg (see Van Paradijs, Kouveliotou \& Wijers (\cite{vanP00})
for a  review).  Frail et  al.  (\cite{Frai01}) have  recently claimed
that considering corrections for possible beaming effects the range of
the inferred high-energy   release is restricted  to  a narrower  band
around $10^{51}$ erg.  Current models invoke the collapse of a massive
star   into   a  black  hole    (Woosley  \cite{Woo93},    Paczy\'nski
\cite{Pacz98}) or the merging  of two compact  objects (e.g.  Lattimer
\& Schramm \cite{Latt74}).  The intrinsic brightness of GRBs and their
afterglows   allows one  to probe  the   nature of their  distant host
galaxies and   potentially the star-formation   history of  the  early
universe (Lamb \& Reichart \cite{Lamb00}).

Since the first  detection of an optical   counterpart to a GRB,  most
searches have been based either  on comparing a  single epoch image to
Digital Sky Survey (DSS) images in order to search  for new objects or
by comparing images taken through the same  filter at different epochs
in order  to find fading,   transient objects.  However,  many optical
transients are  fainter than the DSS  limit  at the  time of the first
optical follow-up observations, and in some  cases the afterglow light
curves show ``plateaus''  that  could disguise their transient  nature
(GRB~000301C;  Masetti  et   al.   \cite{Mase00a}, GRB~001007;  Castro
Cer\'on et     al.  \cite{Cast01})  or     even brightness   increases
(GRB~970508;  Castro-Tirado   et al.  \cite{Cast98}; Pedersen   et al.
\cite{Pede98}).  Hence, it   is clear that  the identification process
would benefit  from alternative identification  techniques.  One  such
alternative  is using colour-colour   selection techniques similar  to
those used for  quasar selection for many  years (e.g.  Warren  et al.
1991 and references therein).  Before  the Beppo-SAX era colour-colour
diagrams were applied to deep late-time CCD images with the purpose of
detecting the  presence  of quiescent objects   with anomalous colours
within GRB error boxes.  When the probability  of finding one of these
objects in a small GRB error  box was low  ($<10^{-3}$) the object was
considered potentially  GRB  related.   Examples of such   objects are
extragalactic (e.g.   quasars  and AGNs;  Vrba  et al.  \cite{Vrba95},
Luginbuhl  et al.  \cite{Lung95}) as  well  as Galactic (e.g.  neutron
stars; Sokolov  et  al.   \cite{Soko95},  flare  stars;   Gorosabel \&
Castro-Tirado  \cite{Goro98}, white dwarfs;  Motch, Hudec \& Christian
\cite{Motc90}, novae; Zharykov, Kopylov \& Sokolov \cite{Zhar95}).

The first reported attempt to exploit colour-colour diagrams for early
afterglow identification of GRBs  was carried out  at the beginning of
the  Beppo-SAX era with optical images  of GRB~970111 taken $19$ hours
after the  trigger  (Gorosabel   et al.   \cite{Goro98a}).    However,
GRB~970111  did not  show any  detectable  optical   emission and  the
feasibility  of   colour-colour discrimination  techniques    was  not
demonstrated.    Recently,  Rhoads  (\cite{Rhoa00a})  has presented  a
detailed discussion of colour-colour discrimination techniques and the
feasibility of using them to distinguish GRB power-law spectral energy
distributions   from curved  (black-body) stellar  spectra.   In an  a
posteriori  analysis,  \v{S}imon et al  (\cite{Simo01}) show afterglow
optical   colours  cluster in  a   precise   position on the   optical
colour-colour   diagrams.  In  this  paper we   present the  result of
implementing such  principles by  applying an  automatic colour-colour
discrimination  software pipe-line to   data taken for GRB~001011 just
$\sim 8 $  hours after the  trigger.   We demonstrate that the  colour
selection techniques can  be   successfully applied to  identify   GRB
afterglows, including the use of near-IR data.

In   Sect.~\ref{Observations} we describe    the observations  of  the
GRB~001011   error-circle.     In  Sect.~\ref{Method}  we    detail  a
candidate-selection   method    based   on     the    study    of  the
optical/near-infrared colours  of   the  objects in the   field.    We
illustrate the technique by describing  its application to GRB~001011,
which enabled the discovery  of its optical/near-infrared counterpart.
Section~\ref{results}  shows  the characteristics   of  the GRB~001011
afterglow      and   its   likely  host       galaxy.     Finally,  in
Sect.~\ref{discussion}  and Sect.~\ref{conclusion} we discuss and list
the conclusions of our work.

\section{Observations}
\label{Observations}

GRB~001011 was detected  on October 11.6631  UT 2000 by both  the Wide
Field  Cameras (WFC) and the  Gamma-Ray Burst  Monitor (GRBM) on board
the Beppo-SAX satellite and localised with an accuracy of $5^{\prime}$
(Gandolfi et al. \cite{Gand00a}).   The position was later  refined to
$2^{\prime}$   prior to  our observations  ($\alpha_{2000}=18^h  23^m
4.32^s$,  $\delta_{2000}=-50^{\circ} 53^{\prime} 56\farcs4$;  Gandolfi
et al.  \cite{Gand00b}).

Optical (R-band) and near-infrared (J  and Ks bands) observations were
carried out  with the 1.54-m  Danish Telescope (1.54D)  and the 3.58-m
New  Technology Telescope  (NTT), both at   ESO, La Silla,  on October
$11.9700$--$12.0361$ UT,  between  $7.37$ and $8.95$   hours after the
burst.  Optical and near-infrared comparison images were obtained with
the same telescopes during   the following week.  Deep R-band   images
were obtained at the 1.54D in April 2001 and at the 8.2-m UT1 of ESO's
Very Large Telescope  (VLT) in May 2001, 6--7  months after the burst.
Table \ref{table1} displays the observing log.

\begin{table*}[t]
\begin{center}
\caption{Journal of observations of the \object{GRB~001011} optical/near-IR counterpart.}
\begin{tabular}{@{}lccccc@{}}
Telescope&Date UT & Seeing & Filter & Exp. Time & Mag.  \\
         &        & (arcsec) &      &  (sec)    &      \\
\hline
NTT(+SOFI, LF) &11.9700--11.9828/10/2000&  1.5& Ks& 15$\times$60 & 17.56$\pm$0.05\\
NTT(+SOFI, LF) &11.9834--11.9959/10/2000&  0.9& J     & 15 $\times$60 & 19.26$\pm$0.05$^{\dag}$\\
NTT(+SOFI, LF) &11.9964--12.0092/10/2000&  0.7& Ks& 15$\times$60 & 17.58$\pm$0.04$^{\dag}$\\
1.54D(+DFOSC)  &11.9934--12.0361/10/2000&  1.2& R     &  5 $\times$600& 20.99$\pm$0.05$^{\dag}$\\
1.54D(+DFOSC)  &14.0029--14.0521/10/2000&  1.0& R     &  5 $\times$600& 23.74$\pm$0.23\\
NTT(+SOFI, SF) &17.9794--18.0138/10/2000&  1.0& Ks& 30$\times$60 & $>$ 21.3$^{\star}$\\
1.54D(+DFOSC)  &19.3710--19.4064/04/2001&  0.9& R     & 10 $\times$200& $>$ 24.3$^{\star}$\\
1.54D(+DFOSC)  &20.3647--20.4242/04/2001&  0.8& R     & 14 $\times$200& $>$ 24.6$^{\star}$\\
1.54D(+DFOSC)  &22.3741--22.4149/04/2001&  1.1& R     & 6  $\times$600& $>$ 24.5$^{\star}$\\
VLT(+FORS1)    &20.1618--20.1905/05/2001&  0.9& R     &  8 $\times$300& 25.38$\pm$0.25  \\
\hline
 $\star$ 3$\sigma$ upper limit. & & & & \\
   \multicolumn{6}{l}{$\dag$  Images used to construct the colour-colour diagram.}\\
\hline
\label{table1}
\end{tabular}
\end{center}
\end{table*}

The near-infrared NTT observations were made with the infrared spectrograph
and imaging camera ``Son OF Isaac'' (SOFI)  using both the Large Field mode
(LF, October 11) and the Small Field mode  (SF, October 17).  The fields of
view  (FOVs) in SF  and LF   modes are $2\farcm   4  \times 2\farcm  4$ and
$4\farcm 9  \times 4\farcm 9$,  respectively.   The 1.54D observations were
made with  the  Danish Faint Object Spectrograph   and Camera (DFOSC) which
provides a FOV  of $13\farcm 7 \times 13\farcm  7$.  Consequently, both the
optical and  near-infrared images taken a few  hours after the  GRB trigger
covered the entire refined WFC error box.  The last set of deep images were
acquired with the  FOcal  Reducer and Spectrograph  (FORS1)  mounted at UT1
(Antu) of the VLT.

\begin{table*}
\begin{center}
\caption{Secondary standards in the field of \object{GRB~001011}}
\begin{tabular}{@{}lcccccccc@{}}
  & RA(J2000)   & Dec(J2000)     & Ks                   & J & R & \\
\hline
1 & 18:23:12.45 &$-$50:53:34.4 & 13.91 $\pm$ 0.03 & 14.33 $\pm$ 0.03 &  15.64  $\pm$ 0.04\\
2 & 18:23:11.34 &$-$50:53:57.7 & 16.47 $\pm$ 0.06 & 17.34 $\pm$ 0.05 &  20.26  $\pm$ 0.06\\
3 & 18:23:03.47 &$-$50:53:41.8 & 15.46 $\pm$ 0.04 & 16.17 $\pm$ 0.04 &  17.59  $\pm$ 0.04\\
4 & 18:23:01.65 &$-$50:54:17.0 & 15.92 $\pm$ 0.04 & 16.71 $\pm$ 0.06 &  18.38  $\pm$ 0.05\\
5 & 18:23:00.72 &$-$50:55:13.3 & 14.90 $\pm$ 0.03 & 15.28 $\pm$ 0.03 &  16.24  $\pm$ 0.04\\
6 & 18:23:00.04 &$-$50:54:36.4 & 14.97 $\pm$ 0.03 & 15.36 $\pm$ 0.03 &  16.33  $\pm$ 0.04\\
\hline
\label{table2}
\end{tabular}
\end{center}
\end{table*}

\subsection{Photometric Calibration}

\begin{figure*}[hbtp]
\begin{center}
% an mbox for alignment
\mbox{
{\includegraphics[height=151mm]{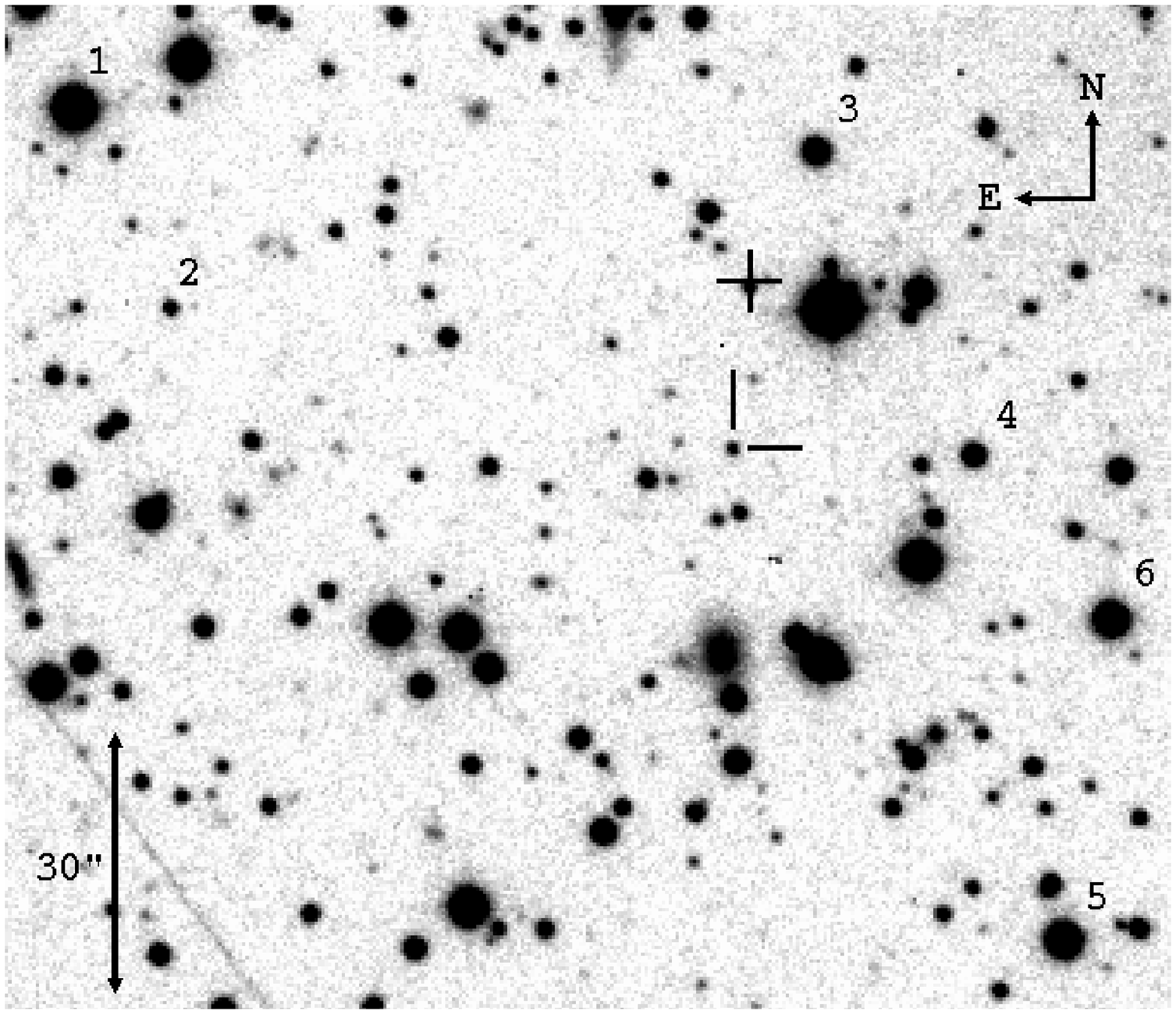}}\quad
}% end of mbox

\vspace{-0.22cm}
\hspace{-0.32cm}
\mbox{
{\includegraphics[height=42.63mm,angle=-90]{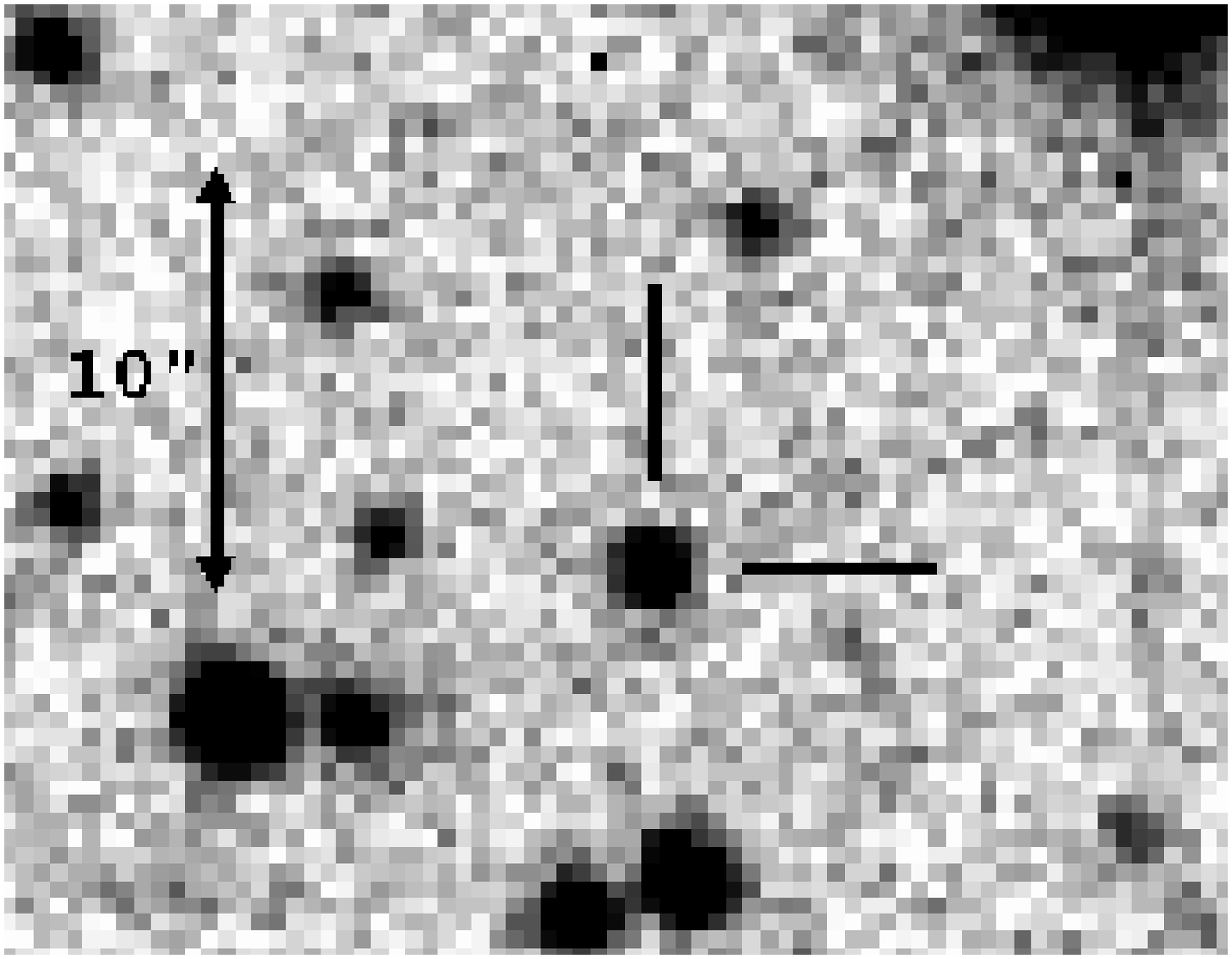}}\quad
{\includegraphics[height=40.63mm,angle=-90]{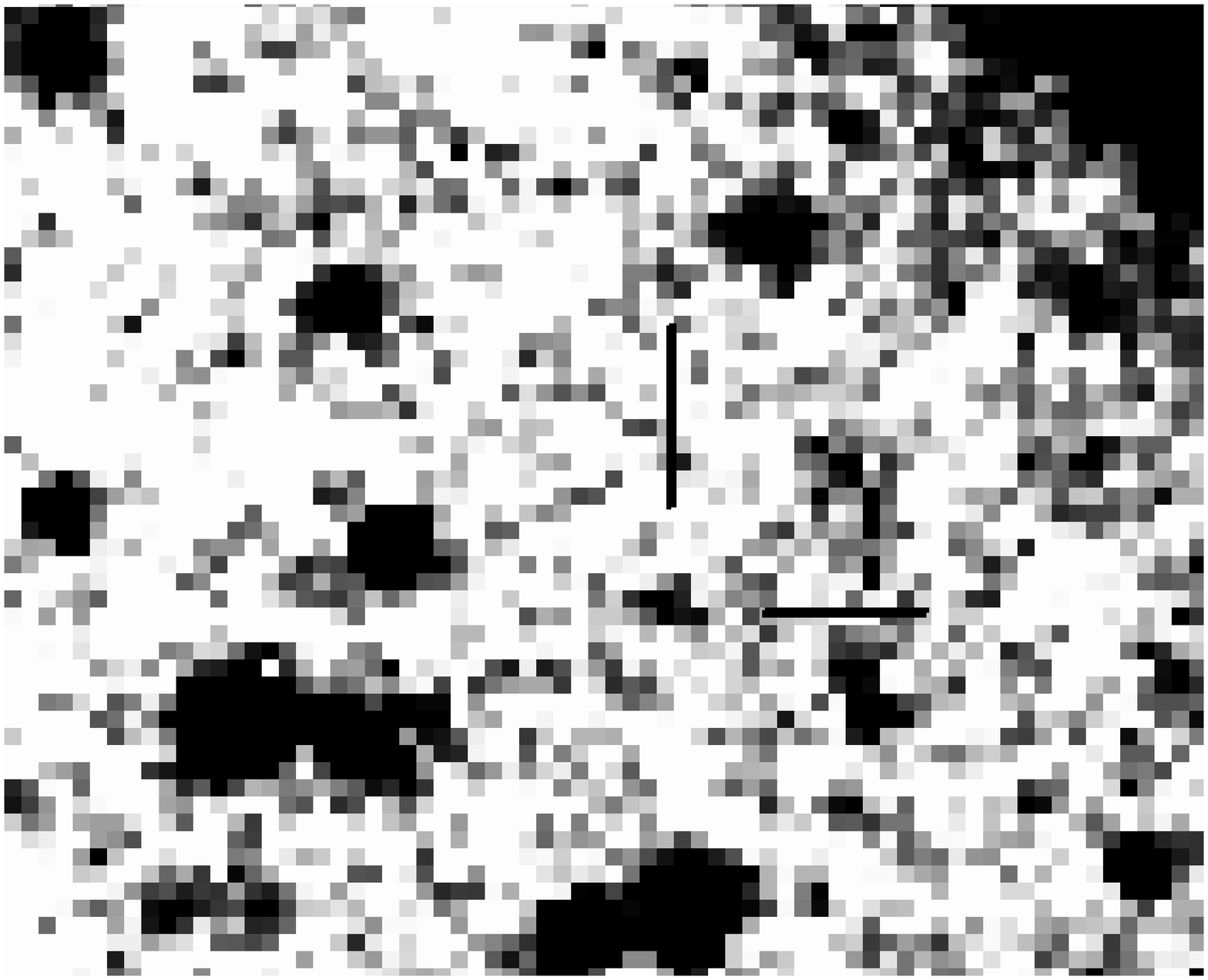}}\quad
{\includegraphics[height=41.63mm,angle=-90]{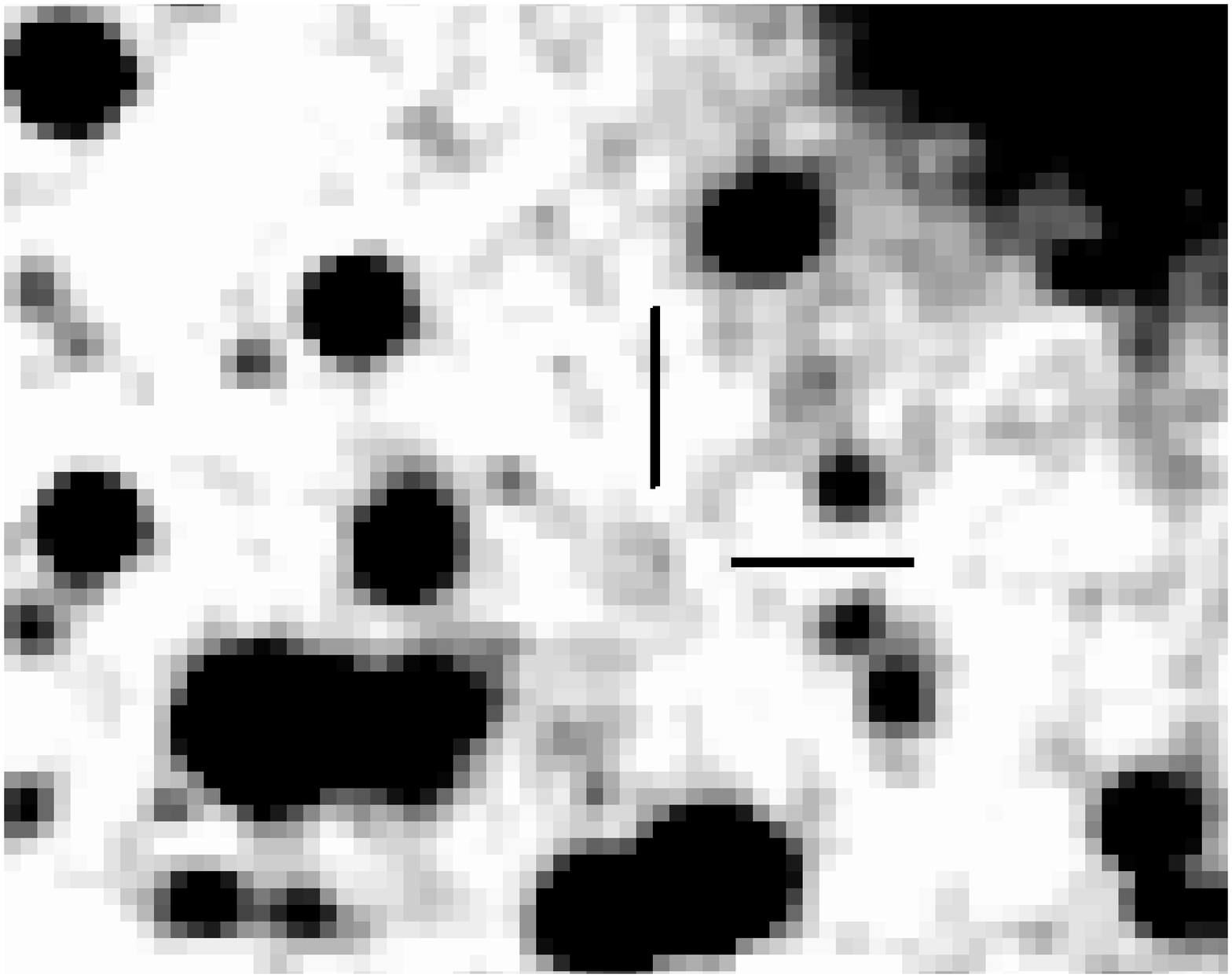}}\quad 
{\includegraphics[height=40.63mm,angle=-90]{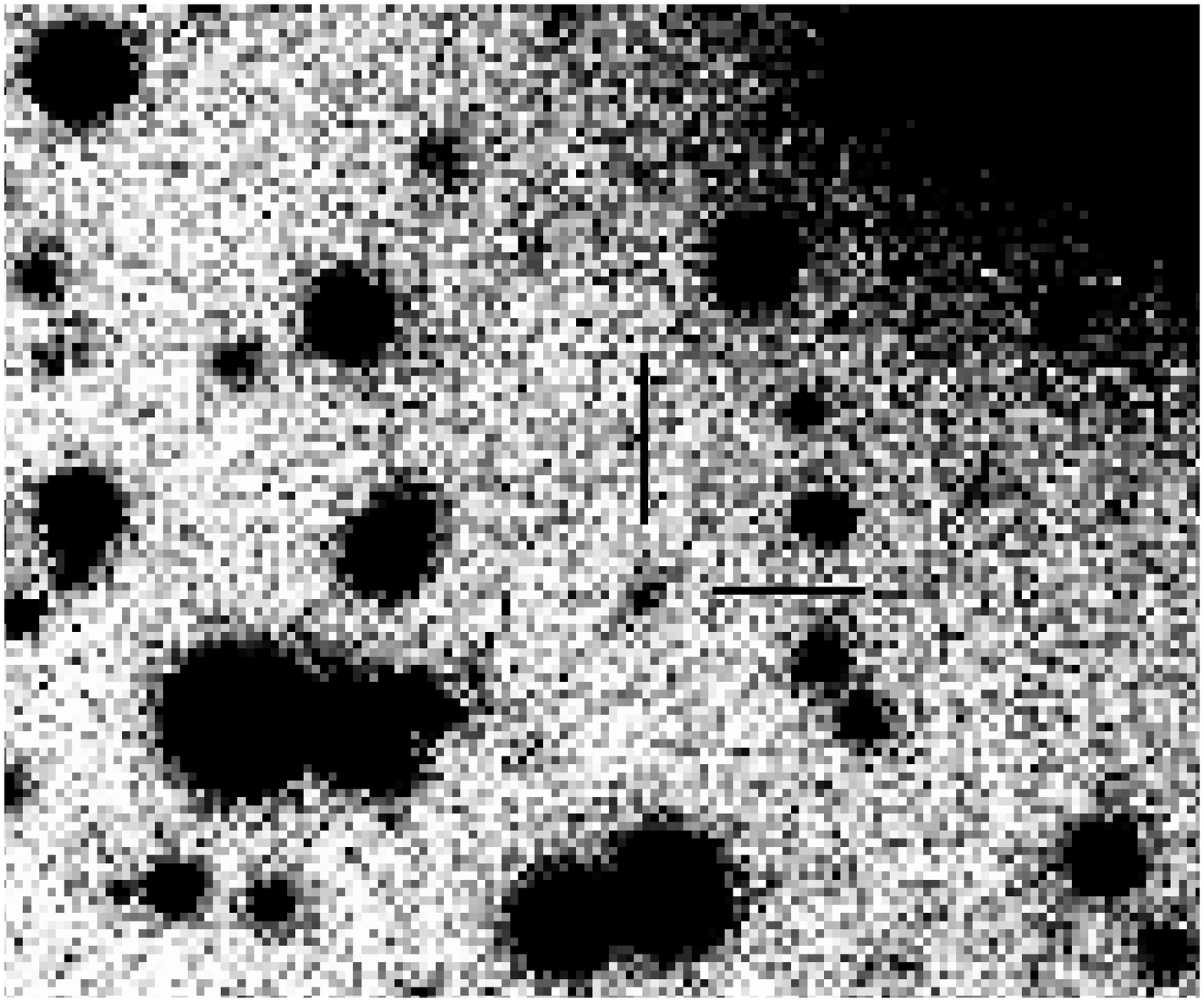}} 

}% end of mbox

\vspace{0.1cm}
\hspace{-0.23cm}
\mbox{
{\includegraphics[height=42.37mm,angle=-90]{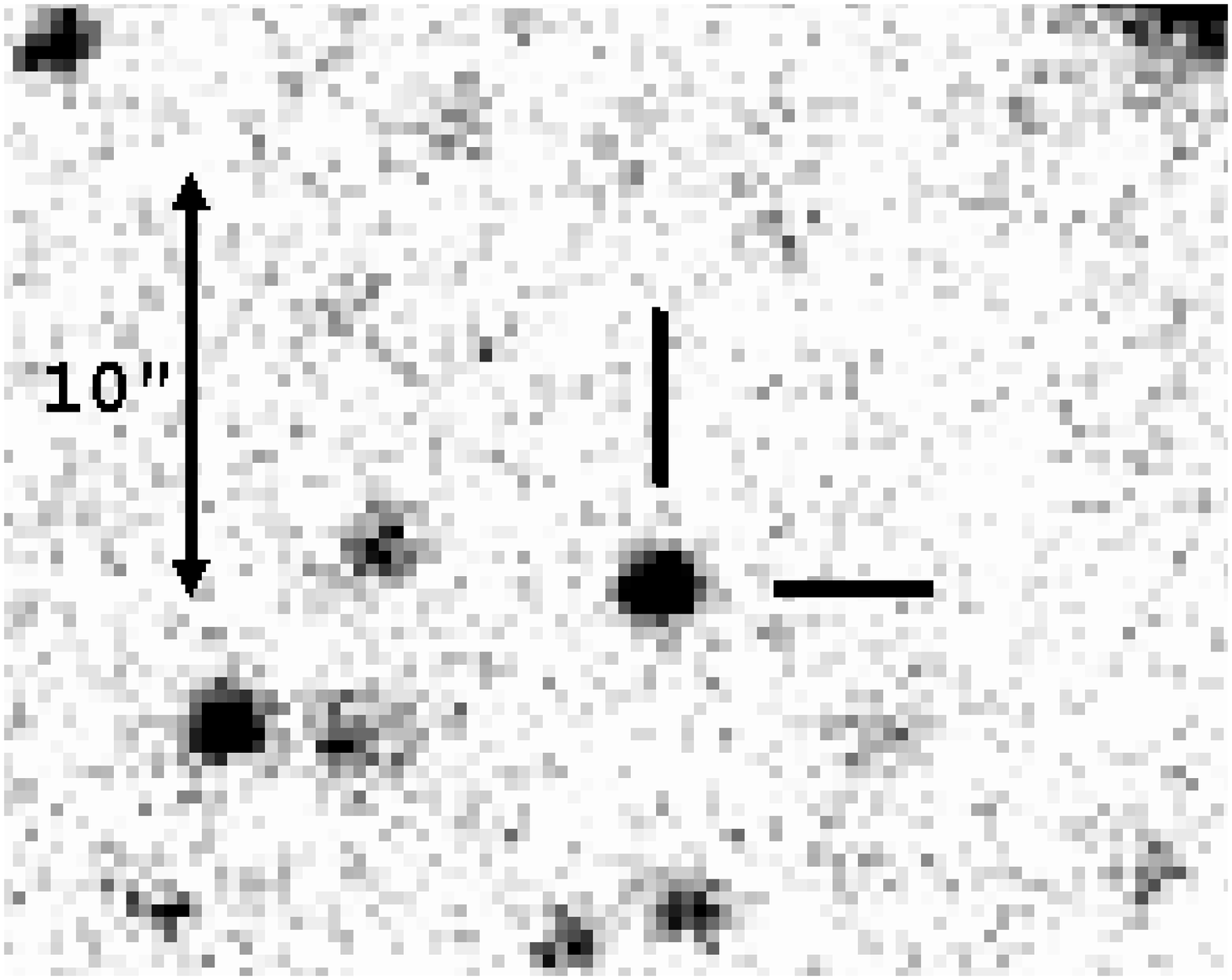}}\quad
\hspace{-0.11cm}
{\includegraphics[height=41.3mm,angle=-90]{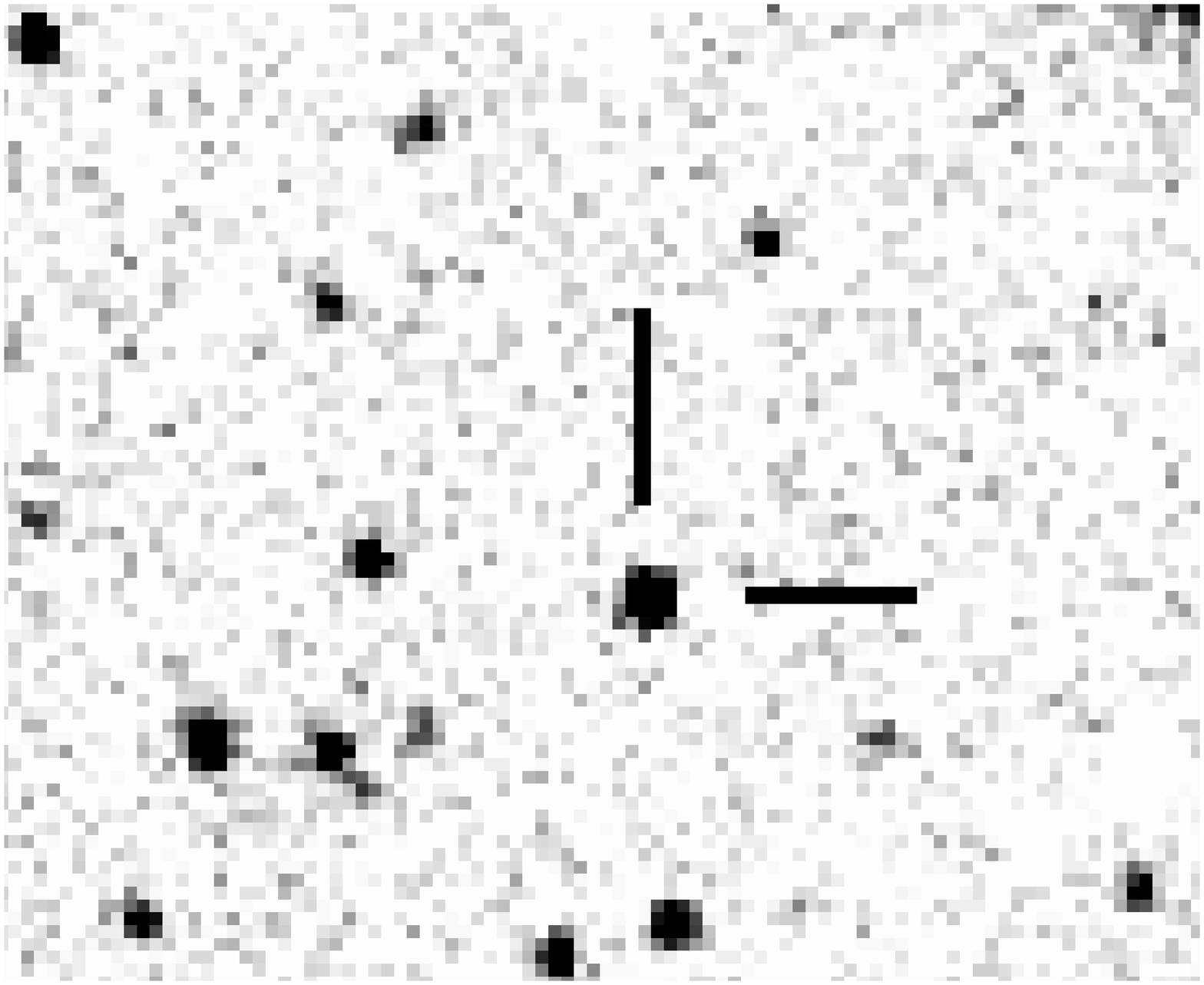}}\quad
\hspace{-0.11cm}
{\includegraphics[height=41.70mm,angle=-90]{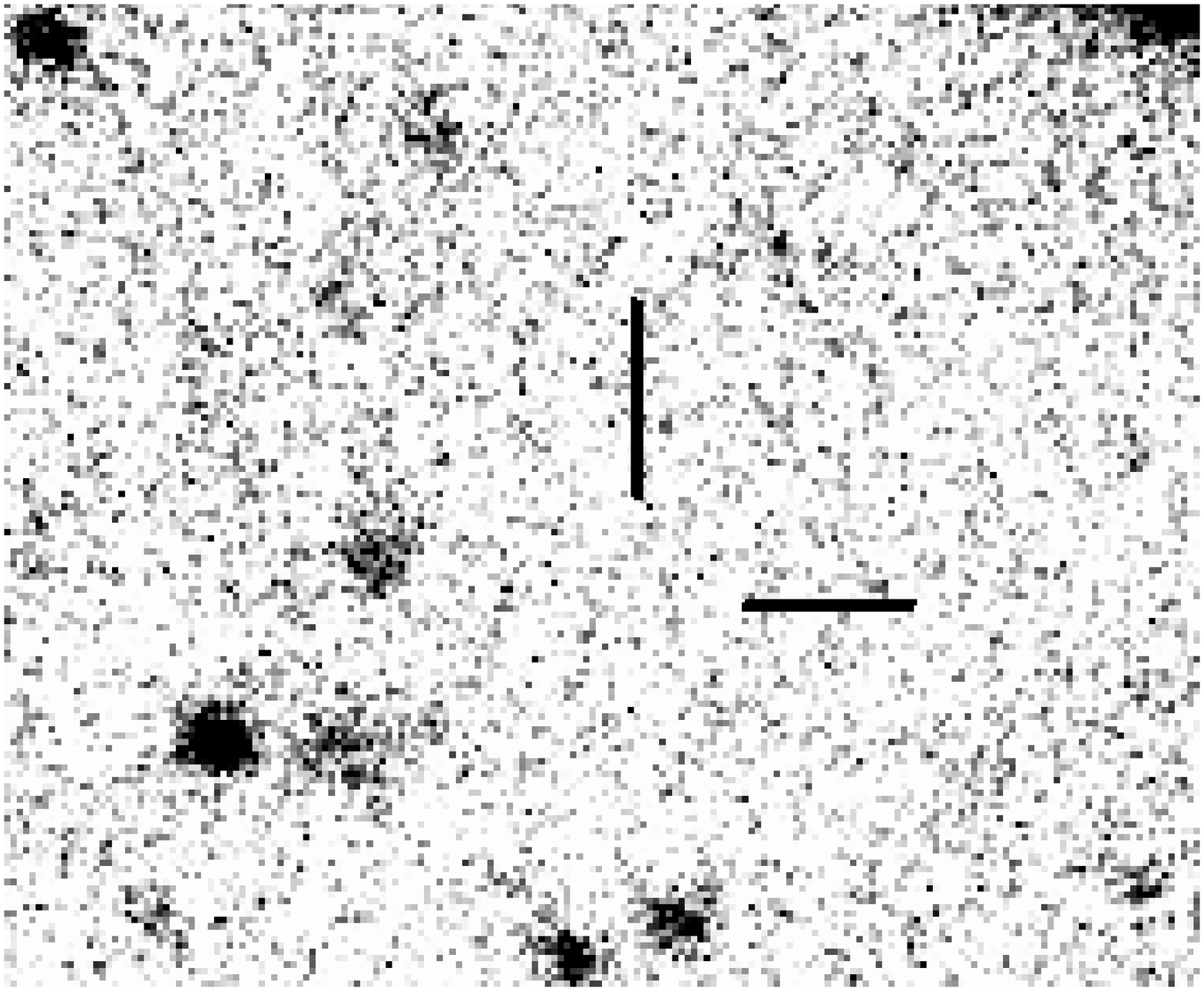}} 
\hspace{0.17cm}
{\includegraphics[height=41.15mm,angle=-90]{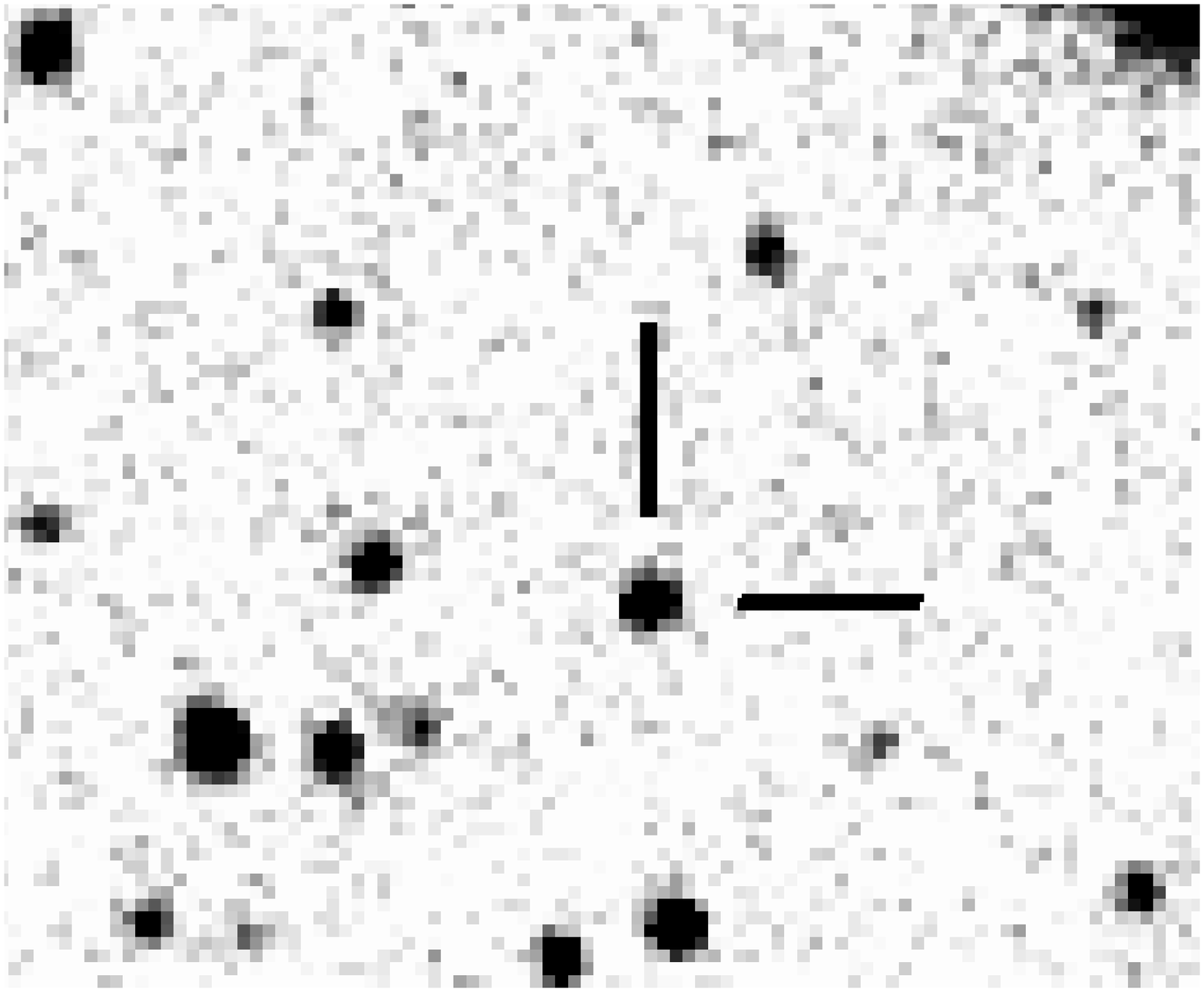}} 
}% end of mbox

\caption {{\em Upper figure:} R-band image of the counterpart (between
tick-marks)   and the  secondary standards  in   the field (see  Table
\ref{table2}).   The cross indicates  the position of  the refined WFC
error circle centre.   {\em Central row of  figures:} from the left to
the right side the R-band decay is  displayed; 11.9934--12.0361 UT Oct
2000, 14.0029--14.0521 UT Oct 2000, 19.3710--22.4149 UT April 2001 and
20.1618--20.1905 UT May 2001 (VLT detection of the host galaxy).  {\em
Lower  row   of figures:}  the first   three  panels show  the Ks-band
brightness  evolution;  11.9700--11.9828  UT, 11.9964--12.0092  UT and
17.9794--18.0138 UT Oct 2000.   The fourth figure displays the  J-band
detection on 11.9834--11.9959 UT Oct 2000.}
\label{images}
\end{center}
\end{figure*}

The field was  calibrated in the  near-infrared using  observations of
the standard stars sj9178, sj9013,  and sj9106 at different airmasses.
These  standards and  thus the  near-IR  measurements reported  in the
present paper are  based on the  JHKs photometric system introduced by
Persson et al.   (\cite{Pers98}).   The derived zero-point  error  was
0.03 mag in both  the J and  Ks bands.  In the  optical the  field was
calibrated using observations of the Landolt field SA 107 at different
airmasses (Landolt  \cite{Land92}).   The   zero-point error  for  the
R-band was estimated  to be 0.02  mag.  In Table~\ref{table2}  we give
the  RJKs magnitudes for 6  secondary standard stars in the GRB~001011
field  (see  Fig.~\ref{images}).  The  photometry  of  these stars was
performed using the DAOPHOT-II package (Stetson 1987, 1997).  The RJKs
zero-point  errors were added in  quadrature  to the measurement error
derived by the DAOPHOT-II package, giving the  error in the magnitudes
shown in Tables \ref{table1}  and  \ref{table2}.  Stars \#1,  \#5, and
\#6 correspond to  the USNO-A2.0 stars U0375-35233918, U0375-35222053,
and U0375-35221378,  which are listed  to   have R-band magnitudes  of
$15.3$, $16.1$, and $16.3$,   respectively.  Using our calibration  we
find  R=$15.64\pm0.04$,  R$=16.24\pm0.04$  and R$=16.33\pm0.04$   (see
Table~\ref{table2}).  Based on these three stars  we have calculated a
mean R-band  offset  of   0.17 mag  with   respect  to  the  USNO-A2.0
catalogue.   Therefore, the calibration used  in  the present paper is
offset by 0.17  mag relative to  the preliminary mean zero-point based
on the USNO-A2.0  catalogue,  which was  used in the  discovery report
(Gorosabel et al. \cite{Goro00a}).

\section{Candidate selection using colour-colour diagrams}
\label{Method}

The main principle behind colour selection of optical afterglows (OAs)
is  that they have power-law spectral  energy distributions, which can
be distinguished from  curved thermal stellar spectra in colour-colour
plots.  This is a  principle that has been  used to select quasars for
many years (Warren et al. \cite{Warr91}  and references therein). This
fact  can be used  to carry out a first  colour-based  selection of OA
candidates.  These candidates can  subsequently  be checked by  second
epoch observations  aimed at detecting  variability.  The advantage of
having an  identification scheme that  is not  based  on the transient
nature of  OAs  is  that it allows   fast   follow-up spectroscopy  or
polarimetry, also for OAs that are fainter than the DSS limit.

All afterglows observed so far as well  as theoretical fireball (e.g.\
Sari, Piran \& Narayan \cite{Sari98};  M{\'e}sz{\'a}ros~\cite{Mesz01};
Piran~\cite{Piran01}) or cannonball  models (Dado, Dar \&  De R\'ujula
\cite{Dado01}) show that  GRB  optical/infrared afterglow  decays  are
reasonably  well described  by  a spectral  index independent  of time
($F_{\nu} \sim \nu^{-\beta}$, $\beta$ not a function  of time).  Thus,
the colours should  remain approximately constant  with  time, and the
technique  is applicable  any  time after the  GRB.   In the case that
$\beta$ is   a  function of  time,  then  the afterglow decay  is  not
achromatic.  Thus, if  the images used  to derive the colours are very
separated in time   and $\beta$ varies  strongly  with time (extremely
unlikely for GRB afterglows), the magnitudes can not be shifted easily
to  the  same  epoch  assuming    a  power  law  decay $F_{\nu}   \sim
t^{-\alpha}$, with an unique achromatic  value of $\alpha$ for all the
bands.   However,   if the  images are  contemporaneous   (or at least
quasi-simultaneous as in the case of  GRB~001011) the error introduced
by  the decay  epoch-scaling factor  is   negligible, even  if $\beta$
depends  on time (see  Sec.~\ref{implement} for colour error estimates
due to the epoch-scaling factor).  So, the technique is not restricted
to   achromatic  afterglow   decays,  if quasi-simultaneous  data  are
used. {\em Therefore,  the method should be valid  at  any epoch after
the gamma-ray burst  at least  until the  emission of  the host galaxy
becomes dominant, usually weeks after the gamma-ray event}.

\begin{figure*}[h]
\begin{center}
\epsfig{file=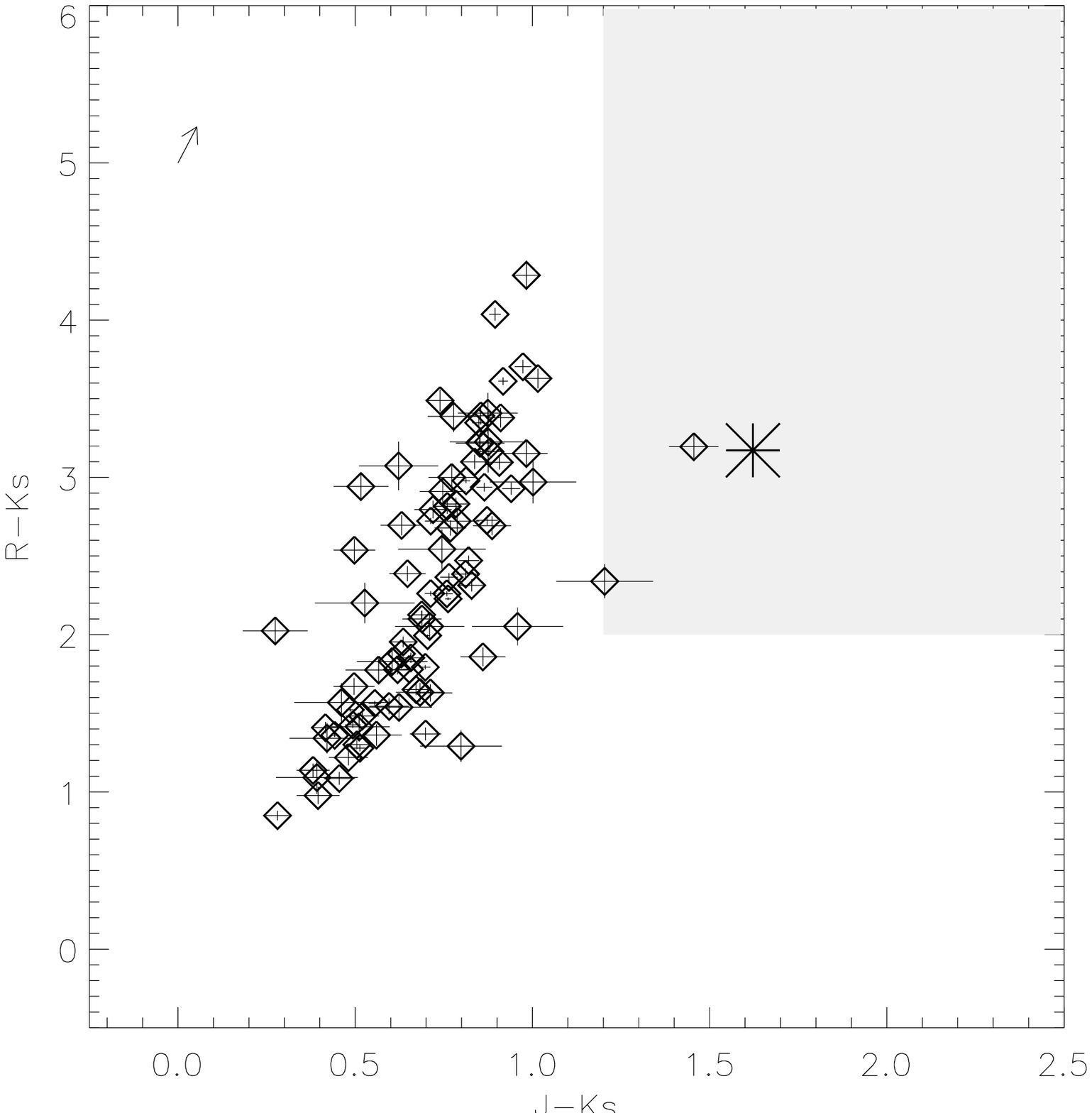,width=8.0cm}
\epsfig{file=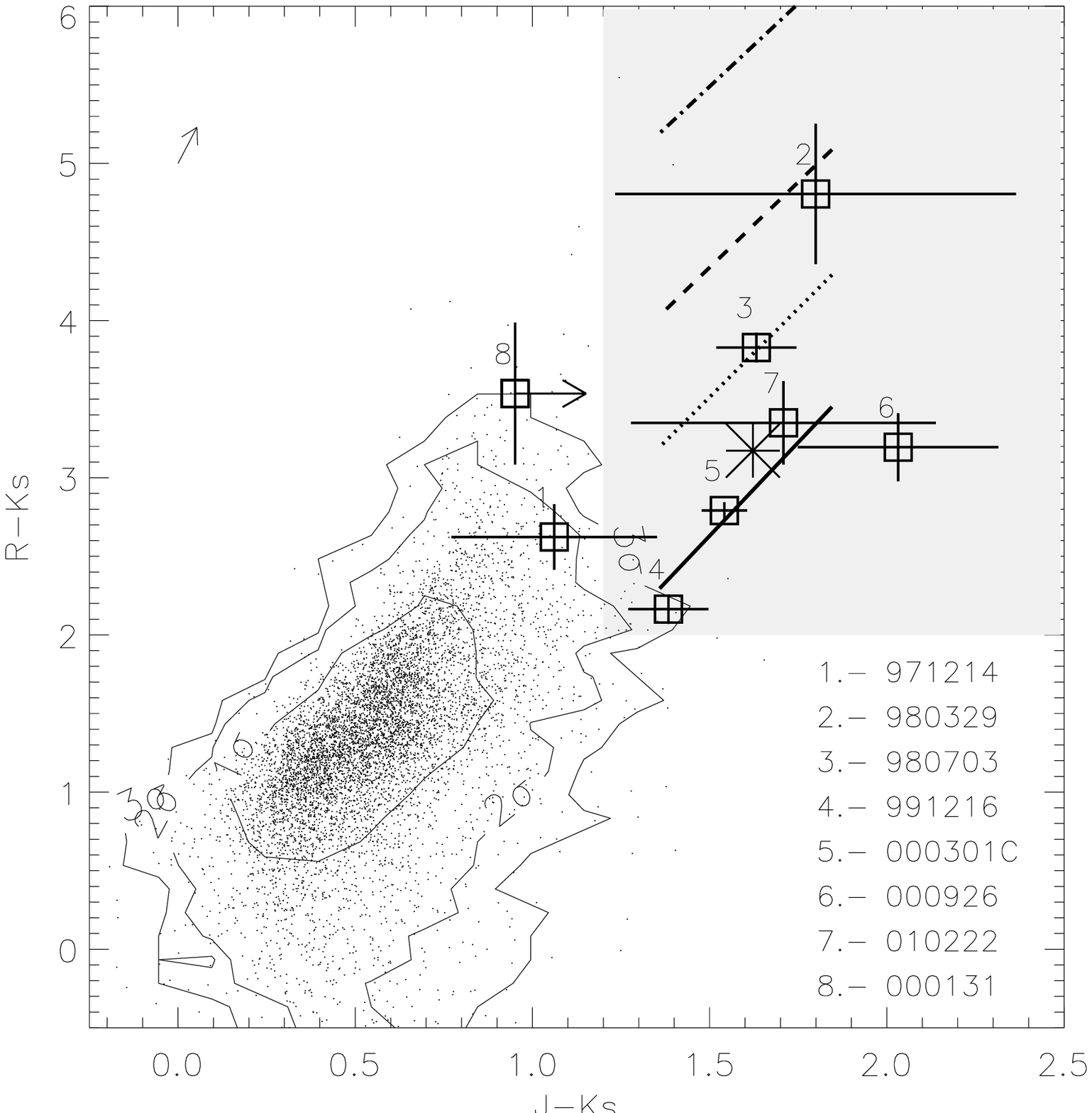,width=8.0cm}
\epsfig{file=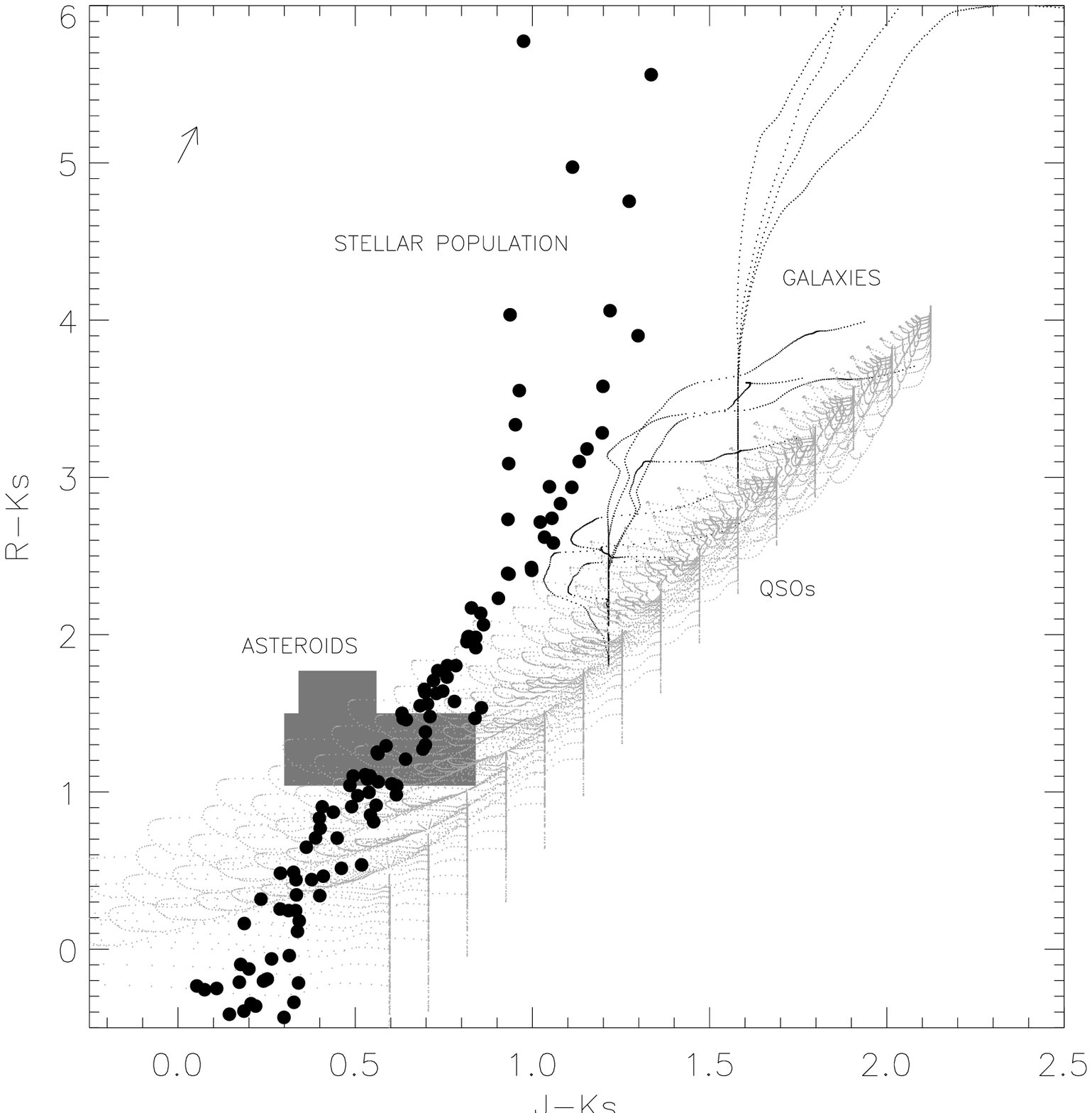,width=8.0cm}
\epsfig{file=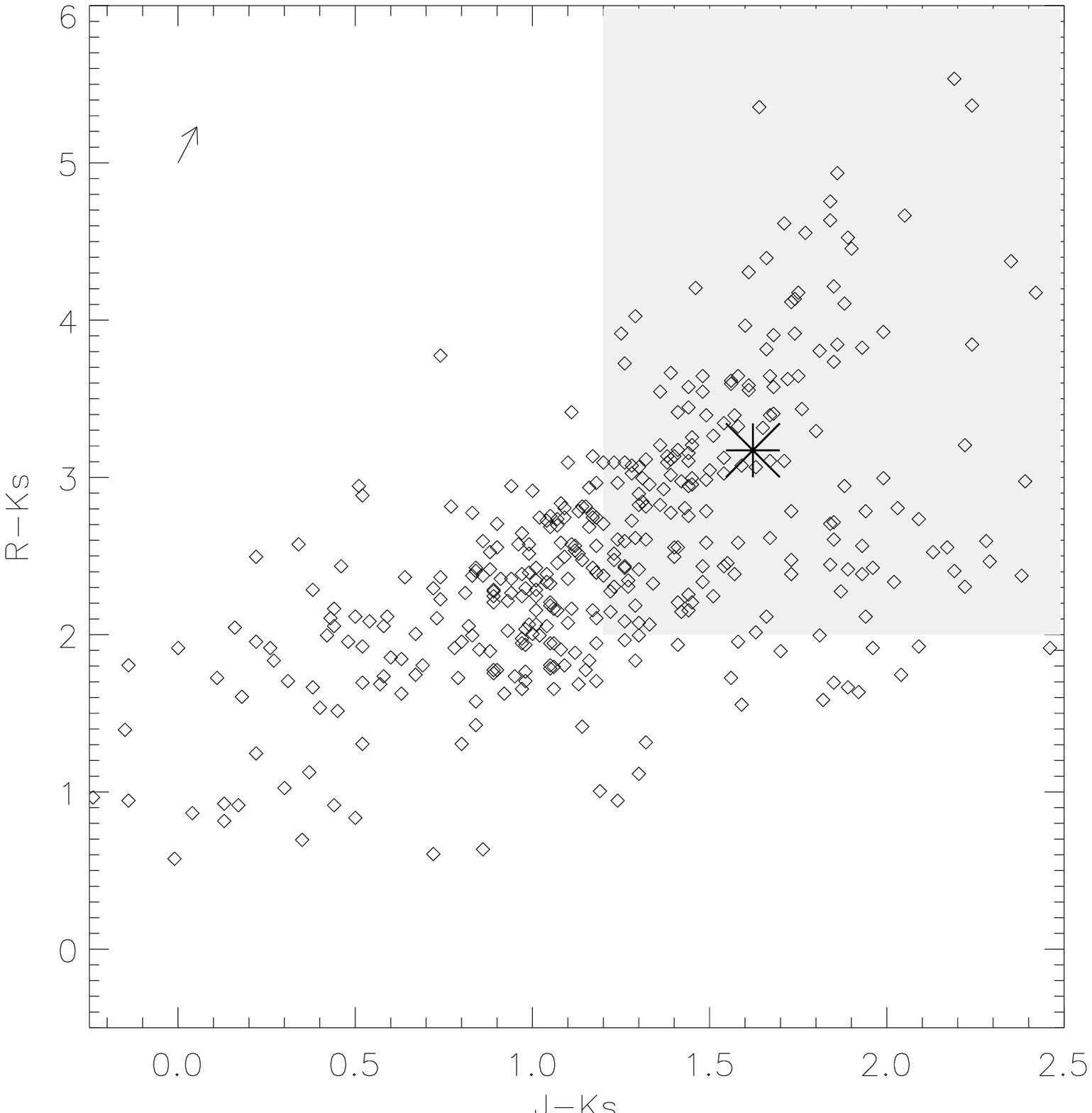,width=8.0cm}
\caption{\label{colour}  {\it Upper left panel:}   The panel shows  an
R$-$Ks versus J$-$Ks colour-colour diagram  for the objects inside the
GRB~001011 error box (open diamonds).  The star represents the colours
of  the GRB~001011  optical/near-IR  counterpart.  The  errors  of the
counterpart colours are smaller than the size of the star.  The shaded
region  shows the  OAs colour-colour  space   locus determined  by our
selection criteria  \rm{R$-$Ks} $> 2.0$,  \rm{J$-$Ks} $> 1.2$.  As can
be seen, the colours of the GRB~001011 counterpart are consistent with
the shaded  area and inconsistent  with almost  all the other  objects
within the error circle.  Only two  non-transient objects in the error
box have colours consistent   with  the shaded  area.  They  are  very
likely quasars  or compact  galaxies (see Sect.~\ref{contami}).   {\it
Upper right panel:} The figure shows an R$-$Ks vs J$-$Ks colour-colour
diagram for   the 10097 2MASS+USNO  sources   (dots) found in  a  $\pm
5^{\circ}$ window around GRB~001011.  As  it is shown, the counterpart
(star) exhibits  colours inconsistent at  least  at a  3$\sigma$ level
with     the    objects  in       the    2MASS+USNO   catalogue   (see
Sect.~\ref{implement}  for  information  on  the  iso-density  contour
levels).  As can be seen, the 2MASS+USNO  catalogue traces most of the
colours shown  in   the upper  left  panel.  The   seven open  squares
labelled from 1 to 7 represent the colours of afterglows with measured
R$-$Ks and J$-$Ks colours to date. The additional square labelled with
number 8 represents GRB~000131 which  was not  detected in the  J-band
(Andersen et al.  2000).  As can be seen,  at least seven of the eight
open squares and  the star are consistent with  the shaded area.   The
eighth open square (GRB~000131) is  likely consistent with this region
too.  The solid straight line inside the  shaded region represents the
colour trace of a  pure power-law SED with  $\beta$ ranging from $0.6$
to  $1.5$.  When Lyman-$\alpha$ blanketing  is taken  into account the
pure power-law SED (straight   solid line) is shifted  upwards  (three
dashed  lines).  From the  bottom to the  top,  the three dashed lines
show the colours of a power-law SED when the Lyman-$\alpha$ blanketing
is   considered for  an  afterglow    at  $z=5.0$, $5.5$   and  $6.0$,
respectively.   As can  be   seen, the higher  the  redshift  the more
distant the power-law SED from the stellar trace  is.  {\it Lower left
panel:} The  figure displays synthetic  colour-colour traces for stars
(filled   circles),  galaxies  (dashed curves)    and  quasars (dotted
curves).   The shaded polygonal background  region represents the loci
occupied  by the  asteroid samples    of  McCheyne, Eaton \&   Meadows
(\cite{McCh85}) and Gaffey  et al.  (\cite{Gaff93}).   As can be seen,
the  bunch of  objects in the  left upper  panel (located outside  the
shaded area) can be reproduced by asteroid, stellar, galaxy and quasar
synthetic colours.  However, the   colours inside the shaded  area can
only  be obtained by galaxy  and  quasar templates, which makes  these
objects a potential source of contamination.  Most of the galaxies can
however  be eliminated  because  of their  non-stellar  profile.  {\it
Lower right  panel:} we have  over-plotted the colours of the galaxies
from the NTT deep  field (Fontana et  al. \cite{Font00}) to illustrate
the    problem  of     the    contamination     by  galaxies      (see
Sect.~\ref{contami}).    About $45\%$  of   the galaxies show  colours
consistent with the ones of OAs (shaded rectangle).  {\it General:} In
the four panels the  length of the arrow and  its orientation show the
magnitude   of   Galactic   extinction  (A$_{\rm  R-Ks}=0.23$, A$_{\rm
J-Ks}=0.05$) and    its   direction   in  the    colour-colour  space,
respectively.  All sources have been dereddened with the corresponding
Galactic  extinction   given   by   Schlegel,   Finkbeiner  \&   Davis
(\cite{Sche98}).}
\end{center}
\end{figure*}

% por si hay quer ponerlo en el caption de la figura 2 del paper.
% ([1] GRB~971214; [2] GRB~980329;  [3] GRB~980703; [4] GRB~991216; [5]
% GRB~000301C; [6] GRB~000926; [7] GRB~010222)

\subsection{Description and implementation of an identification pipe-line}
\label{implement}

Due to the transient  nature of OAs it   is of foremost importance  to
have a  fast identification method once   a GRB localisation  has been
released.  It  is therefore very important to  try to  find an optimal
strategy for  the identification of  the OA.    The most  widely  used
strategy during the last three years has been to obtain an image, most
often  in the R-band,  and then to  look for new  sources by comparing
with DSS images (e.g., GRB~980425, GRB~980519, GRB~990123, GRB~990510,
GRB~000301C, GRB~000926).  In  many cases this method  may  still be a
fast and efficient way to localise the burst, but it is limited to the
brightest OAs in fields of relatively little crowding.

Colour-colour selection is an ideal  complement to the comparison with
DSS images since it    can be done   with one-epoch  observations  and
therefore allows fast follow-up  spectroscopy or polarimetry.  We have
therefore    designed  a   fast reduction pipe-line    that constructs
colour-colour  diagrams of objects in   GRB  error boxes.  Before  the
pipe-line can  be  used we  need  to obtain three  nearly simultaneous
images taken in the R, J and Ks filters. This requires the use of more
than one  telescope and a possibility for  fast data transfer, but for
major modern  observatories such as e.g.   ESO and with modern network
facilities this is not a serious limitation. The three images are then
WCS (World  Coordinate System; e.g.\ Mink 1997)  calibrated and fed to
the pipe-line.  The error introduced in the colours $C$ by considering
quasi-simultaneous images instead of simultaneous ones is of the order
of $\Delta C \sim \alpha \frac{\Delta T}{\overline{T}}$, where $\Delta
T$ is the total time-span covered by the set of images, $\overline{T}$
is the mean delay  of the observation with   respect to the  gamma-ray
event and $\alpha$ is the instant power-law decay slope ($F_{\nu} \sim
t^{-\alpha}$).   In case  the  time-span covered  in the  observations
$\Delta T$ is a  non-negligible fraction of the delay  $\overline{T}$,
then all the measurements should be shifted to the same epoch assuming
a   value of $\alpha$.   In  the case  of GRB~001011 the colour-colour
diagram was  constructed using R,  J, and Ks-band images which overlap
in time (indicated with  a dagger in  Table~\ref{table1}). For a value
of   $\alpha=1.33$  (as   was     found    in the     optical,     see
Sect.~\ref{results}) the errors in the  colours are $\Delta C \lesssim
0.1$ mag.

The pipe-line uses several selection  criteria  or filters to  improve
the identification of candidates.  The   first of these filters is  to
reject  sources outside the GRB     error  box.  Once the    spatially
coincident sources have  been found, aperture photometry  is performed
in the three images and a colour-colour diagram is constructed for all
the  objects inside the  error box.   The colours  of the objects  are
dereddened using the Galactic extinction given by Schlegel, Finkbeiner
\& Davis (\cite{Sche98}) in the direction of the GRB error box centre.
In addition,  the  pipe-line   creates a  list   of  high-redshift  OA
candidates, consisting of those   sources detected in the two  reddest
bands but absent in the blue one.

For a  quick analysis, the discrimination algorithm  can be flagged to
only  use  the relative colour  differences  among objects in  the GRB
error box.  We can derive the two-dimensional probability distribution
of the uncalibrated colours in the field and construct the iso-density
contour levels.  Therefore,   given an uncalibrated candidate  we  can
determine the probability that  this object belongs to  the calculated
probability  distribution, i.e.,  whether  its colours  are typical of
objects in  the error  box  or not.   This relative comparison largely
removes the systematic  effect  of  the  Galactic reddening  and  also
eliminates   the influence  of    a  variable (and   colour dependent)
transparency of the  atmosphere, encountered during  adverse observing
conditions.   Therefore,  the colour-colour discrimination   technique
does not strictly require absolute photometric calibration.  The upper
left  panel   of Fig.~\ref{colour}   shows   an  R$-$Ks versus  J$-$Ks
colour-colour diagram for the objects inside  the GRB~001011 error box
(open diamonds).  As  can  be  seen,  the  colours of   the GRB~001011
counterpart (star) are inconsistent  with almost all the other objects
within the error circle.

A  more careful analysis requires  a rough photometric calibration.  A
good representation  of the distribution of colours  of field stars is
given by the 2MASS and USNO-A2.0 catalogues (hereafter 2MASS+USNO), so
the  pipe-line can be  flagged to use this catalogue  to obtain a well
constrained sequence of  non-transient point   sources near the    GRB
position on  the sky.  In the top  right panel of Fig.~\ref{colour} we
show with dots  and contour levels  the position of 10097 sources from
the 2MASS+USNO database.  These 2MASS+USNO sources were selected to be
inside a   $10^{\circ}   \times 10^{\circ}$  window  centred    on the
GRB~001011 error box coordinates.   The  size of  the window  has been
chosen  to contain $\sim  10^4$ sources,  thus  giving a  large colour
sample for   the construction of the  contour  levels in  a reasonably
short  computing time. In order  to  eliminate the possible  reddening
gradient inside the window  each 2MASS+USNO source has been dereddened
by   its individual reddening, rather   than applying a mean reddening
correction to all sources.

The 2MASS+USNO catalogue trace can be used  to determine a preliminary
colour calibration of the field. The colour  calibration can simply be
done by  matching the median of the  colours of the point-like sources
in the field with the median of  the 2MASS+USNO colours.  By using the
median rather than the  average colour of  the field objects, any bias
introduced by objects with anomalous colours,  which may be present in
the  field, is   minimised.   Only point-like   sources are   used  to
determine the median  of the colours,  so  the effect of the  possible
presence of extended objects (galaxy clusters or nebular structures in
general) is  also  eliminated.  As seen in  the  upper  right panel of
Fig.~\ref{colour}  the  median  of the    colour  for  the  2MASS+USNO
catalogue is located at J$-$Ks=0.54 and R$-$Ks=1.33, respectively. The
standard deviation in  (J$-$Ks) and (R$-$Ks)  is  small, 0.35 mag  and
0.75   mag respectively.  For    a  typical Beppo-SAX  error box  size
($10-100$ arcmin$^2$),  at least 25 objects would   be detected in all
bands.    Therefore,  a   colour zero-point   for  the  field  can  be
established with an accuracy of 0.07  mag in (J$-$Ks)  and 0.15 mag in
(R$-$Ks).  This accuracy  is  fully sufficient for   applying absolute
colour selection criteria.

Most  of the 2MASS+USNO  sources are located  in a sequence stretching
from (J$-$Ks,R$-$Ks)   =   ($0.1$,$0.5$) to ($1$,$3$).     As  we will
describe in Sect.~\ref{delineation} the colours of the majority of the
2MASS+USNO sources can be reproduced by stellar spectral templates.

\subsection{The colours of OAs and the selection criterion}

In the upper right panel of Fig.~\ref{colour} we have over-plotted the
colours of GRB~001011   (marked with a star)  and  the colours of  OAs
(open  squares labelled from 1  to 7) with reported quasi-simultaneous
R,  J, and Ks-band magnitudes, i.e.    GRB~971214 (\#1: Diercks et al.
\cite{Dier98}; Ramaprakash et   al.  \cite{Rama98}), GRB~980329  (\#2:
Fruchter  \cite{Fruc99};  Palazzi  et al.   \cite{Pala98}), GRB~980703
(\#3: Vreeswijk et al.   \cite{Vree99}), GRB~991216 (\#4: Garnavich et
al.  \cite{Garn00}), GRB~000301C (\#5:  Jensen et al.  \cite{Jens01}),
GRB~000926  (\#6: Fynbo et  al.   \cite{Fynb00}) and GRB~010222  (\#7:
Masetti et al.  \cite{Mase01a}).   Each GRB colour has  been corrected
for its Galactic foreground  extinction (Schlegel, Finkbeiner \& Davis
\cite{Sche98}).   GRB~000131 was detected in the  R and Ks-band but in
the J-band  only   an  upper limit  was    reported (Andersen et   al.
\cite{Ande00}).  In  order  to visualise that  the  method is valid at
least  up to $z=4.5$  we have  over-plotted with  an  arrow on an open
square (labelled with an 8) the colour constraints  of GRB~000131.  As
can  be  seen in  Fig.~\ref{colour} the GRB~000131  colours are likely
consistent with  the colours  of the  other  seven OAs (shaded  area).
Most of these OAs have R$-$Ks and  J$-$Ks colours that place them well
away  from the  locus of  the  stellar sequence.  For   a given R$-$Ks
colour the OAs are 0.5--1.0 mag redder in J$-$Ks than stars.  The mean
dereddened colours for  the seven OAs are $\overline{(\rm{R-Ks})}_{\rm
GRB} = 3.3\pm0.9$,  $\overline{(\rm{J-Ks})}_{\rm GRB}=1.6\pm0.3$.  The
positions of  the known   OAs in  Fig.~\ref{colour} do  not  show  any
correlation with their redshift.

The  spectral energy distributions  of GRB afterglows  are fairly well
described by pure power-laws in   the optical to near-infrared  range,
$F_\nu \sim \nu^{-\beta}$. Using synthetic photometry we can therefore
determine an    analytic  expression for the  colour    as function of
$\beta$, which is independent of redshift:

$${\rm R - Ks} = 1.525 + \beta \times 1.284$$
$${\rm J - Ks} = 1.035 + \beta \times 0.540$$
 
In the  right upper panel  of Fig.~\ref{colour} we have over-plotted a
solid line    that corresponds to a    pure  afterglow Spectral Energy
Distribution  (SED)  with $\beta$  ranging  from  $0.6$  to  $1.5$, as
expected  for afterglows.    All  the traces  and  objects  (including
afterglows) plotted in Fig.~\ref{colour} have been dereddened by their
corresponding  Galactic   extinction  (Schlegel,  Finkbeiner  \& Davis
\cite{Sche98}).  As  seen, the observed  OAs  are consistent  with the
predicted colours for power-law SEDs.
 
\begin{figure}
\epsfig{file=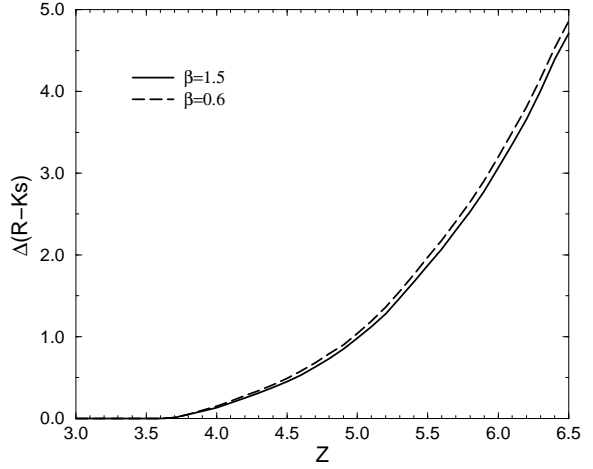,width=\hsize}
\caption{\label{blanketing}   The  figure shows   the reddening of the
  R$-$Ks colour of a pure power-law as a  function of the redshift due
  to Lyman-$\alpha$  blanketing  along   the line  of   sight  to  the
  observer.    The blanketing effects   for the two   extreme cases of
  $\beta=0.6$ (long-dashed   line) and $\beta=1.5$  (solid  line)  are
  plotted.  Most GRB  afterglows are intermediate between these cases.
  The reddening   of the (R-Ks)  colour   due  to  the  Lyman-$\alpha$
  blanketing starts to be noticeable for redshifts $z>3.7$ (see text).
  For $z<3.7$ the blanketing effect is absent.  }
\end{figure}

When  $z\gtrsim4$ the effect   of  Lyman-$\alpha$  blanketing  becomes
noticeable in the R$-$Ks colour, while the J$-$Ks colour is unaffected
until $z\sim8$.  We have  modelled the effect on  the R$-$Ks colour by
imposing Lyman-$\alpha$ blanketing  on  a power-law SED  with spectral
slopes in  the range $0.6  \le \beta \le  1.5$ (M{\o}ller  \& Jakobsen
\cite{Moll90};   Madau  \cite{Mada95}).   The resulting  spectrum  was
integrated  for a range of redshifts,  using the transmission function
of an  R-band filter  as a weight   function.  The result is  shown in
Fig.~\ref{blanketing}.  The effect of Lyman-$\alpha$ blanketing in the
R-band has its onset at $z=3.7$  and is essentially independent of the
intrinsic  spectral slope.   At   a redshift  of   $z \sim  6.5$,  the
attenuation  reaches $5$ magnitudes, implying  that GRB afterglows are
unlikely to  be detected in  the  R-band at redshifts  much beyond $6$
(see  Lamb   \& Reichart   \cite{Lamb00}  for  a  discussion    on the
detectability of very high redshift GRBs).

The effect of the Lyman-$\alpha$ blanketing on the colour-colour space
can  be visualised on the upper  right panel of Fig.~\ref{colour}.  As
the  Lyman-$\alpha$   forest   shifts   into  the  R-band  sensitivity
wavelength  region   the value of R$-$Ks   increases  moving the solid
straight-line (pure power-law) upwards (dotted, dashed and dash-dotted
line).  For simplicity we  have only over-plotted  the  loci of a pure
power-law  affected  by blanketing when  the afterglow  is  located at
$z=5.0$  (dotted line),  $5.5$  (dashed  line) and $6.0$  (dash-dotted
line).   As  can  be  seen in   Fig.~\ref{colour} the   effect of  the
blanketing is to further  separate the colours expected  for power-law
SEDs from the non-transient objects.

In order  to establish a list of  OA  candidates we use  the following
colour selection criteria:  \rm{R$-$Ks} $> 2.0$, \rm{J$-$Ks} $>  1.2$.
These   criteria      define   the    rectangular  shaded      area of
Fig.~\ref{colour}.   This   area  contains the  expected   colours for
power-law OAs  as well as OA SEDs  modified  by Ly-$\alpha$ blanketing
and at the  same time has  very little contamination from  the stellar
sequence.

\subsection{Location of other astrophysical objects in the diagram}
\label{delineation}

In  order to improve  the discrimination method, the different regions
which the non-transient classes of objects occupy in the colour-colour
diagram  must  be determined.  To  achieve  this  goal we  either  use
synthetic  photometry with    spectral templates from   the literature
(stars, galaxies and quasars,  following Wolf, Meisenheimer \& R\"oser
\cite{Wolf01a}, Wolf et al. \cite{Wolf01b}) or use colour observations
from the literature (asteroids).

For stars we calculate R$-$Ks and J$-$Ks colours based on the spectral
atlas of Pickles   (\cite{Pick98}), which  contains spectra  of  $131$
stars with spectral types ranging from O5 to  M8.  It covers different
luminosity  classes, but concentrates on main  sequence  stars, and it
also contains  some spectra  for high  metalicities.   The colours for
these $131$  stars are  shown  with filled circles   in the lower left
panel of Fig.~\ref{colour}.

The expected colours   for a population  of galaxies   were calculated
using the galaxy spectral templates by Kinney et al.  (\cite{Kinn96}).
This library consists of ten SEDs  averaged from integrated spectra of
local galaxies ranging in wavelength from $125$ nm to  $1000$ nm.  The
input  spectra of  quiescent    galaxies were  sorted    by morphology
beforehand into four templates  and the starburst galaxies were sorted
by  colour into six groups yielding  six more templates.  Colours were
calculated for all individual spectra  redshifted from $0$ to $2.0$ in
steps of $0.01$. The derived galaxy colours are displayed in the lower
left panel of Fig.~\ref{colour} with dashed lines.

The quasar library  is designed as  a three-component model: We  add a
power-law continuum    with an emission-line   contour based   on  the
template  spectrum by Francis et  al.  (\cite{Fran91}).  The intensity
of  the emission-line contour was varied  only globally, i.e.  with no
intensity dispersion   among the lines.    The slope of  the power-law
continuum  $F_\nu  \sim \nu^{-\beta}$   was varied  in $15$  steps  of
$\Delta \beta  = 0.2$ ranging from  $\beta  = -0.8$ to $\beta  = 2.0$.
The  library was  calculated for $151$   redshifts ranging in steps of
$\Delta z = 0.02$ from $z=0$ to $z=3$.

The number of quasars with  redshift $z>3$ having R$\lesssim$23.5 (the
limiting magnitude of  the discovery DFOSC  image) expected inside the
refined WFC  error  box   is    low  ($\sim10^{-2}$; based on      the
contamination  estimates  given   in  Sec.~\ref{contami} and   in  the
fraction of $z>3$ quasars  present  in the $10^{th}$ recompilation  by
V\'eron-Cetty  \&  V\'eron \cite{Vero01}), therefore  synthetic quasar
colours have not been constructed for  redshifts $z>3$.  The synthetic
quasar   colours   are   represented    with   a   dotted   line    in
the lower left panel of Fig.~\ref{colour}.

For  satellites looking for GRBs  in  or near  the ecliptic  plane, such as
HETE-II,  it  is  also important    to consider  the contamination  due  to
asteroids, especially  as these objects  do not  appear in  the sky surveys
that  are usually  used   (e.g.   DSS).    Based on   the  asteroid  colour
recompilation by  McCheyne, Eaton \& Meadows   (\cite{McCh85}) we derived a
mean    colour     of    $\overline{(\rm{R-Ks})}=1.49   \pm   0.28$     and
$\overline{(\rm{J-Ks})}=0.45   \pm  0.11$  based   on   a sample  of  seven
asteroids.  An independent sample of albedos  of nine asteroids reported by
Gaffey et al.  (\cite{Gaff93}) gives $\overline{(\rm{R-Ks})}=1.27 \pm 0.23$
$\overline{(\rm{J-Ks})}=0.57  \pm 0.27$, consistent with  the first sample. 
Thus, the location of  asteroids in the  (J$-$Ks) vs (R$-$Ks) colour-colour
diagram is far from the region where  we find GRB afterglows, ensuring that
the colour-colour  technique is  effectively screening out  asteroids.  The
loci of the above  mentioned two asteroid  samples  is shown by  the shaded
background region of the lower left panel of Fig.~\ref{colour}.

\subsection{Rejection of non-transient sources with similar colours as OAs}
\label{contami}

Although there are many stars  in the field, the stellar contamination
is low.  As can be seen in  the upper right panel of Fig.~\ref{colour}
the GRB colour-colour space region (shaded rectangular area) is almost
completely  beyond  the 3$\sigma$  contour  level   of the  2MASS+USNO
catalogue.

The overlap  with  galaxy colours is   more significant.  According to
Huang  et al.  (\cite{Huang01}) we would  expect $\sim 4 \times 10^3$,
$10^4$ and  $2.4 \times 10^4$ galaxies  per square degree with R$<$21,
R$<$22 and  R$<$23, respectively.  However, in  the  $21 \lesssim$ R $
\lesssim 23$ magnitude  range  only $\sim  20\%$  of the  galaxies are
compact under   observing  conditions of  $1\farcs0$ (Calar  Alto Deep
Imaging Survey, Wolf et al.  \cite{Wolf01b}).   Thus, we have designed
a selection   filter to permit identification  and  elimination of all
extended objects  from the colour-colour  plot.  To apply this filter,
the Full Width Half Maximum (FWHM) distribution for all objects in the
field is derived.   Next, the mode  of the distribution is calculated,
and  all the objects  beyond $\pm 3\sigma$ from  the  FWHM mode can be
rejected.   Based on  the NTT   deep  field catalogue  (Fontana et al.
\cite{Font00}) we estimate that $\sim 45  \%$ of compact galaxies have
colours consistent  with GRBs   (see     the lower right  panel     of
Fig.~\ref{colour}).   Taking into account    these estimates we  would
expect $\sim$ $1$, $3$ and $8$ compact  galaxies brighter than R$=21$,
R$=22$ and R$=23$, respectively, in the 2 arcmin radius WFC error box.
Thus, we can eliminate most of the galaxy contamination especially for
bright OAs.  This  demonstrates  that, besides  the colour, {\it   the
magnitude  of the source is  an important parameter that together with
the colour shows the unusual nature of a  source}.  However, we do not
wish to exclude faint candidates since one of the important quests for
future   searches is to  extend  the detection  sensitivity to fainter
magnitudes (Fynbo et al.  2001b).

Quasars  are not  so easy to  eliminate  because  they are  point-like
sources and their SEDs can be described by power-laws similar to those
of afterglows.  Fortunately (in   this  context), quasars  are   rare.
According to  Boyle,   Jones \&  Shanks   (\cite{Boyl91}), Hartwick \&
Schade (\cite{Hart90}) and Wolf et al.   (\cite{Wolf99}) the number of
quasars per square degree with R$<21$, R$<22$ and R$<23$ is $\sim 80$,
$200$ and  $400$, respectively.   The   number of quasars  inside  the
GRB~001011 error  box would then be $0.3$,  $0.7$ and $1.4$, depending
on the depth of the image  (R$<21$, R$<22$ and R$<23$).  As mentioned,
GRB afterglows exhibit spectral indices  $\beta$ ranging from $0.6$ to
$1.5$.  Quasars  instead show values of  $\beta$ between $0$  and $1$.
Hence,  afterglows  tend to be redder   than  quasars.  Therefore, the
quasar contamination  is mainly due to  the reddest quasars.  Only one
third  of the  quasars  show indices  $\beta >   0.6$  (Francis et al.
\cite{Fran91}).  The pipe-line  is able to  exploit this property  and
has  an option  to reject  objects  with spectral indices smaller than
$0.6$.   The  spectral    index of each    object  is  calculated   by
least-square  fitting a   power-law SED to    the  R, J,   and Ks-band
magnitudes.

Afterglows which are reddened by Lyman-$\alpha$ blanketing are shifted
into a region of the colour-colour diagram which is occupied by L-type
brown dwarfs.  Although  the density  of brown dwarfs  is  low, of the
order of one per square degree (Kirkpatrick et al.  \cite{Kirk99}; Fan
et  al.  \cite{Fan00}) down to a  Ks-band  limiting magnitude of $20$,
this  complicates the   secure  colour-colour  identification of  high
redshift afterglows in large error boxes.

Therefore, we consider that for a  first selection of candidates the impact
of contamination  by non-transient sources is  not severe, at least for GRB
error boxes smaller  than $\sim10$ arcmin$^2$ and  possibly up to $\sim100$
arcmin$^2$.

\section{Results}
\label{results}

\subsection{Detection of the optical and near-infrared afterglow of
GRB~001011}

In the   upper  left panel   of Fig.~\ref{colour}  we show   with open
diamonds the R$-$Ks and J$-$Ks colours for all objects detected in all
three bands in the  GRB~001011 error box.   Most of the sources lie in
the sequence also seen for the 2MASS+USNO sources.  When the selection
method described above was applied, six  candidates were selected, but
three of them were rejected because of  their non-stellar shape (using
the selection filter described in Sect.~\ref{contami}).

\begin{figure}[t]
\begin{center}
  {\includegraphics[height=8.0cm]{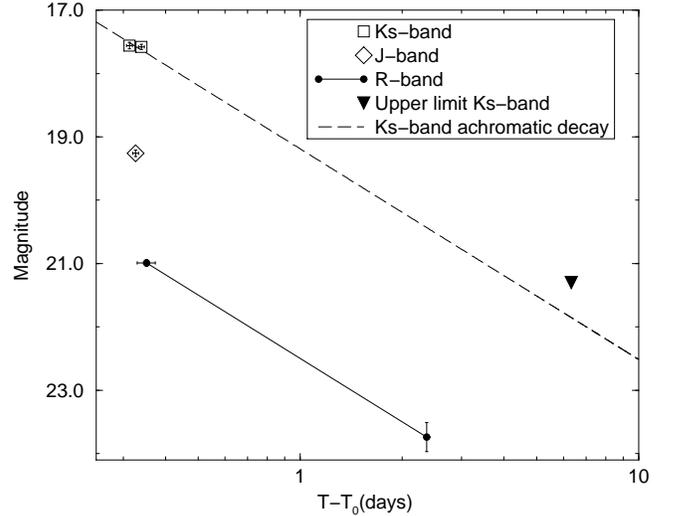}}
\caption{\label{lightcurve} The figure shows  the afterglow history in
  the  Ks, J, and  R-bands.   The two  open  squares show the  Ks-band
  detections and the diamond the J-band magnitude.  The symbol size of
  these  points   has  been enlarged   in order  to  make   easier the
  visualisation of the error bars contained in the symbols.  The solid
  line represents the $\alpha_{R}=1.33$ decay  between our two  R-band
  measurements (filled circles).  The triangle shows the Ks-band upper
  limit imposed  on   $17.9794$--$18.0138$ UT  Oct  2000   by  the NTT
  observations.   The error bars along   the horizontal axis represent
  the exposure  time of the observations.   As can be seen  the second
  Ks-band    observation, the first  R-band detection   and the J-band
  measurement overlap  in time, so  they  have been  the basis of  the
  colour-colour   plot  presented  in    Fig.~\ref{colour}   (see  the
  observations  indicated with  a dagger  in Table~\ref{table1}).  The
  dashed line shows that  the NTT upper  limit  is consistent with  an
  achromatic decay ($\alpha_{\rm R}=\alpha_{\rm Ks}$).}
\end{center}
\end{figure}

Second epoch observations  performed on Oct $14.0029$--$14.0521$ UT at
the 1.54D  showed   that  one of  the   three candidates  had   almost
disappeared (see  Fig.  \ref{images}).  The  remaining two sources are
compact, non-transient  and have colours  that are consistent with the
OA colour-colour space  locus.  Both objects can be  seen in the upper
left  panel of Fig.~\ref{colour} as  the  two closest open diamonds to
the   GRB~001011 colours (star) and  consistent  with the shaded area.
The magnitudes of the two objects  are R=$21.04$ and R=$21.38$, so the
contamination is   consistent  with the  expected number   of  compact
galaxies and quasars with 21$<$R$<$22 estimated in Sect.~\ref{contami}
(1+0.3  $\lesssim$ number of   compact galaxies  +  quasars $\lesssim$
3+0.7).  The colours  of both objects can  be  reproduced by synthetic
traces  of quasars or  galaxies    (see Fig.~\ref{colour} lower   left
panel), so they are likely quasars or compact galaxies.

An astrometric solution based  on 50 USNO  A2-0 reference stars yields
$\alpha_{2000}=    18^h  23^m  04.56^s$,$    \delta_{2000}=-50^{\circ}
54^{\prime} 15\farcs8$ (uncertainty 1\farcs0) for the OA, indicated in
Fig.~\ref{images}.   These  coordinates  correspond  to  the  Galactic
coordinates $l=343.686^{\circ}$,  $b=-16.535^{\circ}$  and a  Galactic
colour excess  E(B$-$V)=0.10     (Schlegel,   Finkbeiner   \&    Davis
\cite{Sche98}).

The position of the counterpart is  fully consistent with the position
of  the refined $2^{\prime}$ WFC  radius error circle (Gandolfi et al.
\cite{Gand00b}) and  is located only $19  \farcs  5$ from  its centre.
The counterpart turned  out to be the object  with  the reddest J$-$Ks
colour among the  90 point-like sources  located within the  GRB error
box and   detected  in the  three bands.    The  dereddened colours of
GRB~001011   are    (J$-$Ks)$_{\rm     GRB~001011}=      1.62\pm0.06$,
(R$-$Ks)$_{\rm   GRB~001011}=3.17\pm0.07$,  fully consistent  with the
dereddened colours of GRBs detected to date in the  R, J, and Ks-bands
and close to the  trace of OA  power-law SEDs (see solid straight line
of  Fig.~\ref{colour}  right upper  panel).     The  counterpart  was
point-like in the   optical and  in   the near-IR, showing the    same
point-spread function as neighbouring stars.   In order to improve the
photometry given by the automatic  pipe-line more accurate  photometry
was carried out  by means of the  DAOPHOT-II photometry package.   The
final photometry of the counterpart is shown in Table~\ref{table1} and
displayed in Fig.  \ref{lightcurve}.  The  decay indexes in the R  and
Ks    bands are   $\alpha_{\rm R}   =1.33  \pm  0.11$  (solid  line of
Fig.~\ref{lightcurve})  and   $\alpha_{\rm   Ks}  =  0.24  \pm  0.73$,
respectively.   The large  error in the  determination of $\alpha_{\rm
Ks}$ is due to the short time-span between the two Ks-band detections.

In order to measure a possible deviation  from an achromatic decay, we have
calculated  the  expected magnitude  difference between  the second and the
first   epoch  Ks-band detections  if  $\alpha_{\rm  R}=\alpha_{\rm Ks}$ is
assumed.  The   prediction ($\Delta$Ks  =  0.12  mag)  is only 1.5~$\sigma$
different from  the measured magnitude  difference ($\Delta$Ks$ =  0.02 \pm
0.06$), so an achromatic decay  is an acceptable  approximation for our two
Ks-band detections.  A power-law fit to  the two Ks-band detections, fixing
an achromatic decay index ($\alpha_{\rm Ks}=\alpha_{\rm R} =1.33 \pm 0.11$,
see dashed  line of Fig.~\ref{lightcurve}), yields  Ks$=21.85 \pm  0.35$ on
$17.9794$--$18.0138$ UT  Oct 2000.  Thus,  the NTT upper limit (Ks $>21.3$,
represented with a triangle in Fig.~\ref{lightcurve}) is consistent with an
achromatic fading. We conclude that our measurements are consistent with an
achromatic decay,  although  we can  not  exclude  a more complex   Ks-band
lightcurve, with plateau   phases  (e.g.  GRB~000301C, Rhoads   \& Fruchter
\cite{Rhoa01a};   GRB~010222,     Masetti   et  al.      \cite{Mase01a}) or
re-brightenings (e.g.  GRB~971214; Gorosabel et al.  \cite{Goro98b}).

Once the  RJKs-band  quasi-simultaneous  magnitudes   (see  magnitudes
indicated with a dagger in Table~\ref{table1}) have been dereddened by
Galactic  extinction  in the direction   of GRB~001011 (E(B$-$V)=0.10;
Schlegel, Finkbeiner \&   Davis  \cite{Sche98}) and shifted  in   time
(assuming an achromatic fading  with $\alpha=1.33$) to the mean  epoch
of the  second Ks-band detection  ($12.0028$  UT), a least-squares fit
provides  a  spectral index of $\beta=  1.25  \pm 0.05$.  The value of
$\beta$  has  been  derived  without considering intrinsic extinction,
which is unknown  for GRB~001011.  Therefore, $\beta=  1.25$ has to be
considered as an upper limit to the actual OA spectral index.

Given that we have measured $\alpha$ and constrained  the value of $\beta$,
these values can be compared to the predictions  given by several afterglow
models.  However,  given  the  poor  coverage of the  light  curve  and the
absence of an estimate of the intrinsic extinction we stress that only weak
conclusions on  the afterglow physical  properties can be deduced  from our
data. If we assume an unextincted afterglow then the upper limit on $\beta$
would be  close to the actual spectral  index.  According  to the adiabatic
expansion of a spherical afterglow in the  slow-cooling regime (Sari, Piran
\&  Narayan  \cite{Sari98}), $\alpha$=3$\beta$/2  (for $\nu  <  \nu_c$) and
$\alpha$=(3$\beta$-1)/2 (for $\nu    >  \nu_c$) is  expected.   Thus,   the
prediction of $\alpha$  given for $\nu < \nu_c$  is $\alpha=1.88 \pm 0.08$,
which is  $4\sigma$ away from  the measured  value.  However, the predicted
value given  for $\nu > \nu_c$ is  $\alpha=1.38 \pm  0.08$ fully consistent
with the measured  value of $\alpha= 1.33  \pm 0.11$.  Besides, the derived
value of  the electron  power-law index, $p=2  \beta  =2.5 \pm 0.1$, is  in
agreement  with those obtained  for other afterglows.  For the more general
and common  case of an extincted afterglow  the  value of $\beta$  would be
decreased and  it could  be compatible  with the  predictions given by  the
spherical adiabatic expansion for the case  of $\nu < \nu_c$.  The measured
value  of $\beta=1.25 \pm 0.05$  is also explainable  in the context of the
cannonball model (Dado, Dar \& De R\'ujula \cite{Dado01}).

\subsection{The host galaxy}

\begin{figure}
\epsfig{file=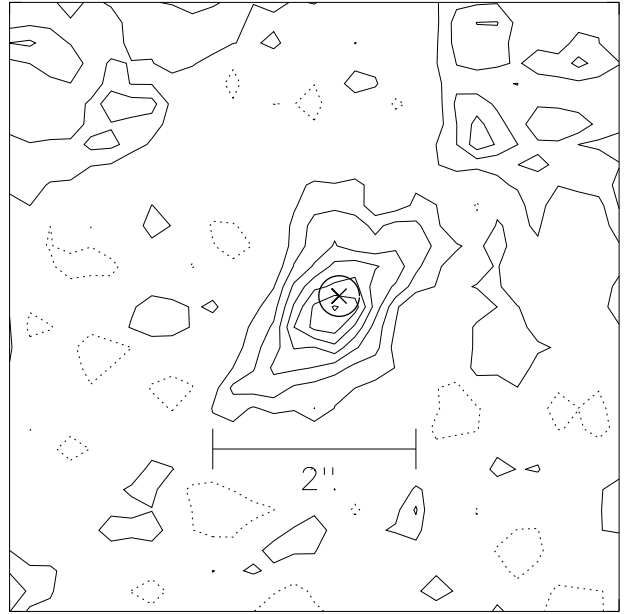,width=9.0cm}
\caption{\label{host} The figure shows a contour  plot of the co-added
  R-band images taken  in May 2001  with the VLT.  The total  exposure
  time is 2400s with a mean seeing of  0\farcs9.  The cross shows the
  position  of the   optical  counterpart, fully  consistent with  the
  object.     The   object is  clearly    elongated  in the North-West
  direction.   The  image has been smoothed  with  a  3$\times$3 pixel
  boxcar filter.  North is to the top and East to the left.}
\end{figure}

When  all the R-band images   taken in April  2001  at the 1.54D  were
co-added, the presence of a very  faint extended object $\sim 3\sigma$
above  the   background   was noticeable.     Further  deep    optical
observations carried out one month   later with the VLT confirmed  the
presence  of an  object    (detected at 8$\sigma$   significance) with
R$=25.38 \pm 0.25$ measured in  a 3\farcs0 diameter circular aperture.
The coordinates of the object  are $\alpha_{2000}=18^h 23^m  04.58^s$,
$\delta_{2000}=-50^{\circ}    54^{\prime}   16\farcs0$    (uncertainty
0\farcs5),  which  is 0\farcs28 from  the OA  position.   Thus, the OA
seems to be  centred on the  extended object.  This strongly  suggests
that the object  is the host galaxy  of GRB~001011.  A contour plot of
the  object   and    the  position  of   the    OA  can   be seen   in
Fig.~\ref{host}. The host galaxy is elongated in the North West--South
East direction (PA = $-$45$^\circ$).

%the best astrometry for the host is 18:23:04.578,-50:54:15.99

\section{Discussion}
\label{discussion}

The  colour-colour discrimination  pipe-line  presented in  this paper
makes use  of the R, J,  and  Ks-band magnitudes.   However, it can be
applied using other bands as well. The best  choice of the bands would
be  those where most  of the curved stellar spectra  are as distant as
possible from the afterglow  power-law SEDs.  A  detailed study of the
optimal  configuration of near-IR  + optical filters is however beyond
the scope of the present paper.  The method can also be generalised to
a  higher number  of filters which   would allow  one to  construct an
N-dimensional colour-colour space.

%For instance,  if we had an  additional detection  in another band we
%could   construct a three  dimensional  colour-colour space, making it
%possible to discriminate between sources having the same projection in
%the J$-$Ks vs R$-$Ks plane.

Obviously, the  colour-colour selection is  only useful for  GRBs that
have  optical  and/or  near-IR  afterglows.  However,  with  8m  class
telescopes the  technique can  easily be used  for bursts as  faint as
R $=  24$.  Hopefully this  technique  will  therefore facilitate  the
detection of optical and/or  near-IR afterglows from a larger fraction
of  well  localised GRBs  than  the  $\approx30\%$  during the  latest
$3$--$4$ years (Fynbo et al. \cite{Fynb01}).

%For  obvious reasons the  colour-colour  selection method technique is
%only  valid  for   GRBs that  do  show  optical  and/or near-IR
%counterparts.   Thus, the so-called dark  GRBs can  not be detected by
%this technique.

Another possible  bias  could affect the   selection of high  redshift
afterglows.  As long as the Lyman-$\alpha$ forest  does not enter into
the bluest observed band the  method  does not introduce any  redshift
selection  bias.   At a  redshift $z  \gtrsim  3.7$ the Lyman-$\alpha$
forest enters the R-band, and the spectrum is no longer well described
by  a single power-law SED.  However,  the  effect would only make the
OAs  redder in R$-$Ks and thereby  move the OAs  further away from the
non-transient sources.

Therefore, as long  as  the object is   detected in the  R  band $\sim
 5\sigma$ above the sky background,  the blanketing does not represent
 any    inconvenience to   locate and   distinguish   objects in   the
 colour-colour  diagram.  Conversely, it would  help to discriminate a
 candidate from the  non-transient sources in   the GRB error  box, at
 least up to a  given upper limit in  the redshift.  For higher values
 of  the redshift where  the blanketing  effect  makes that the R-band
 detection is below $5\sigma$, then the large photometric errors would
 be a difficulty to  distinguish the OA from the  rest of the objects.
 In this  regime the blanketing  starts to represent an inconvenience,
 rather than   an  advantage.  For even   larger   redshift values the
 blanketing effect can be so  severe that the  afterglow is likely  to
 remain undetected in the R-band.

 In the  case   of GRB~001011 the  R-band   detection  is  about $2.5$
magnitudes above the detection limit of the image, hence we could have
tolerated    a blanketing  effect    up  to  $z   \lesssim  5.8$  (see
Fig.~\ref{blanketing}).  We estimate that  for the set of three images
(1.54D R-band,  NTT  J and Ks-band)  used  in the present  study,  the
Lyman-$\alpha$ blanketing helps the discrimination of the afterglow in
the redshift range $3.7 < z <  5.3$.  For a  redshift range $5.3 < z <
5.8$  the GRB~001011 R-band  detection would be below 5$\sigma$ having
R-band photometric errors above $\sim 20$\%, so  the efficiency of the
pipe-line   goes down.  For redshifts $z   \gtrsim 5.8$ the blanketing
would make the   GRB~001011   afterglow undetectable  for   our R-band
images. For other limiting magnitudes and combinations of filters, the
corresponding redshift ranges will be different.

Obviously, the  former discussion is  only dealing  with the impact of
the blanketing on the colours of GRB~001011, and it has not considered
the monochromatic   cosmological dimming factor  given by $D_L(z)^{-2}
\times (1 + z)^{{\alpha}-{\beta}-1}$, where $D_L(z)$ is the luminosity
distance. This effect would enlarge the photometric errors of the high
redshift afterglows  and  hence complicate  their  discrimination (see
Lamb \& Reichart \cite{Lamb00} for a detailed discussion on the effect
of the monochromatic cosmological dimming factor).

The application of the  method to very  reddened afterglows  which are
bright in the  near-IR but with no detectable  emission in the optical
is possible as long as a constraining lower limit on the R$-$Ks colour
can be derived.  This could happen under conditions of high extinction
or with extremely high redshift bursts having $5.8 \lesssim z \lesssim
8$. The   pipe-line considers the potential  case  of highly extincted
and/or redshifted afterglows creating a  list of sources only detected
in the near-IR (both J and Ks bands), and hence  absent in the optical
image.

The efficiency  of the pipe-line  in the $5.8   \lesssim z \lesssim 8$
redshift range depends on the constraint  imposed by the R-band image.
As can be  seen in Fig.~\ref{colour} for  an R-band limiting magnitude
of R $>$ Ks$+2$ a combined detection in both the J  and Ks bands would
be able   to distinguish   the   shaded area  from  the rest    of the
colour-colour space.  The discrimination of undetected optical sources
in  the  R $<$  Ks$+2$ region   is uncertain.   For $z\gtrsim   8$ the
Lyman-$\alpha$  blanketing also   affects   the J-band and   a  single
analysis becomes much more complicated.

Finally, there is a bias against OAs situated in bright host galaxies,
both because the flux of the underlying galaxy may move the integrated
flux of OA + galaxy  away from the shaded  region in Fig. \ref{colour}
and because the galaxy will make the OA look extended.

\section{Conclusion}
\label{conclusion}

We present  a colour based selection  pipe-line of  OA candidates that
only  requires   three  quasi-contemporaneous   images.   The   colour
selection  software  is designed to be  used  in  parallel with, or as
input  to,  the   normal technique,  searching   for transient sources
through the comparison  of observations from (at  least) two epochs or
with the  DSS.  Furthermore,  it   is a   method  that allows   a  fast
identification of candidates  for follow-up spectroscopy also  for OAs
that are   fainter  than   the  DSS  limit   at   the time   of  first
optical/near-infrared observations.   Thus, the technique is not meant
to replace the normal procedure, but to complement it. The combination
of colour-colour and variability  information could be a very powerful
mean   of doing  automated  OA   discovery,  with great  potential for
forthcoming missions like Swift.

The technique has several advantages; it can be applied any time after
the gamma-ray event,   using  a single  set   of images in   different
filters,  and it does     not strictly require  absolute   photometric
calibration.  Another   additional  advantage is  that  the  method is
independent of the redshift, at  least for redshifts with a negligible
Lyman-$\alpha$ blanketing along the line of sight.

In the case that the discrimination method  is based on  the R, J, and
Ks magnitudes,  it becomes independent of  the redshift for $z\lesssim
3.7$.  For    redshifts $3.7\lesssim z    \lesssim 5.3$  the  possible
blanketing     effect  along  the  line of      sight helps the colour
discrimination.   In  the range   $5.3  \lesssim z  \lesssim 5.8$  the
blanketing introduces considerable errors in the R-band photometry and
it starts to be an inconvenient.  For $5.8  \lesssim z \lesssim 8$ the
efficiency is still valid as long as a constraining lower limit on the
R$-$Ks colour  can be derived    (R $>$ Ks$+$2).   For extremely  high
redshift  afterglows ($z\gtrsim 8$)  the  colour-colour discrimination
technique is uncertain.

This kind of analysis is very suitable for  small GRB error boxes such
as the ones reported by  the NFI  of Beppo-SAX  or  for the ones  that
HETE-II  is expected eventually to determine,  in which case the field
contamination of quasars and compact galaxies is small.

Using   this method  we  discovered  the   GRB~001011 afterglow.   The
GRB~001011  optical afterglow  evolution   is consistent with a  decay
index   of $\alpha_{R}=1.33 \pm 0.11$.   With  no  corrections for the
intrinsic  absorption, we   derived  a  spectral  index   of $\beta  =
1.25\pm0.05$.  Therefore, this  value of $\beta$  has to be considered
as an upper limit to the  unextincted afterglow spectral index.  If we
assume negligible  intrinsic absorption,  the  values of  $\alpha$ and
$\beta$  are  consistent with  a  spherical  afterglow model  with  an
electron energy index $p=2.5$.  These values  also would indicate that
the cooling break, $\nu_c$, was located  at frequencies lower than the
R-band $\sim8$ hours after the gamma-ray  event.  This would make from
GRB~001011 a  very interesting system,  since in most cases $\nu_c$ is
higher than the optical frequencies at early times.

Images taken 7 months after the  burst reveal an elongated object with
R$=25.38 \pm 0.25$  fully consistent with  the OA position, likely the
host galaxy of GRB~001011.

\section*{Acknowledgements}
J.  Gorosabel acknowledges support from  the ESO visitors program  and
also the receipt of  a Marie  Curie  Research Grant from  the European
Commission.   We acknowledge the   availability of the 2MASS  and USNO
catalogues.  This  work  was supported by   the Danish Natural Science
Research Council  (SNF).  We are very grateful  to I.~J.  Danziger for
helpful   comments.   We  thank   B.   Montesinos  and D.   Barrado  y
Navascu\'es for fruitful discussions  on  the contamination  by  brown
dwarfs.  The observations presented in  this paper were obtained under
the   ESO Large Program  165.H--0464.    We appreciate  the useful and
helpful comments of the referee, Dr. G. Grant Williams.

\end{document}